\documentclass[a4paper,11pt]{article}
\pdfoutput=1

\usepackage{jheppub}

\renewcommand\thesection{\arabic{section}}

\usepackage[english]{babel}
\usepackage{amsmath,amssymb,amsfonts,commath}
\usepackage{caption}
\usepackage{graphicx}
\usepackage{booktabs}
\usepackage{hyperref}
\usepackage{color}
\usepackage{tikz}
\usepackage{cleveref}

\usepackage{booktabs}
\usepackage{comment}
\usepackage{breqn}
\DeclareMathOperator{\Tr}{Tr} 
\usepackage{appendix}
\usepackage{tabularx}
\usepackage{adjustbox}
\usepackage{orcidlink}

\usepackage{soul}
\setstcolor{red}

\newcommand{\Pf}{{\rm Pf}}

\renewcommand{\Re}{\text{Re}}
\renewcommand{\Im}{\text{Im}}

\title{AMSB in $Sp(N_c)$ Gauge Theories}

\author[a, b]{Digvijay Roy Varier\,\orcidlink{0000-0002-4695-7428}}
\emailAdd{digvijayroyvarier@berkeley.edu}

\author[a, c, e]{Zijian Gu}
\emailAdd{zijiangu@vt.edu}

\author[a, b]{Bea Noether\,\orcidlink{0000-0002-2947-3210}}
\emailAdd{bea\_noether@berkeley.edu}

\author[a,b,d,1]{Hitoshi Murayama\,\orcidlink{0000-0001-5769-9471}}
\emailAdd{hitoshi@berkeley.edu, hitoshi.murayama@ipmu.jp}
\note{Hamamatsu Professor}
\affiliation[a]{Leinweber Institute for Theoretical Physics, University of California, Berkeley, CA 94720, USA}

\affiliation[b]{Ernest Orlando Lawrence Berkeley National Laboratory, Berkeley, CA 94720, USA}

\affiliation[c]{Department of Physics,  Virginia Tech, Blacksburg, VA 24061, USA}

\affiliation[d]{Kavli Institute for the Physics and Mathematics of the
  Universe (WPI), University of Tokyo,
  Kashiwa 277-8583, Japan}

\affiliation[e]{School of Physics, Peking University, Beijing 100871, China}

\abstract{
We present a careful study of the chiral symmetry breaking minima and other potential minima in supersymmetric symplectic QCD ($Sp(N_c)$ with $N_f$ flavors) perturbed by Anomaly Mediated Supersymmetry Breaking (AMSB). Although the case of $N_f = N_c +1$ requires particular care due to the inherently strongly coupled nature of the quantum modified moduli space, we are able to show that all $Sp(N_c)$ theories to which AMSB can be applied ($N_f < 3(N_c + 1)$) possess stable chiral symmetry breaking
minima, which are plausibly continuously connected to the vacua of QCD-like $Sp(N_c)$ theories for large SUSY breaking,
and are protected from runaways to incalculable minima.
}

\begin{document}

\maketitle

\section{Introduction}
The challenge of establishing the phase structure of Quantum Chromodynamics (QCD), corresponding to the observed color confinement with
chiral symmetry breaking, might be tackled \cite{Murayama:2021xfj} by adding small SUSY breaking perturbations to the exact results and phase structures of the supersymmetric versions of these theories (SQCD) \cite{Seiberg:1994bz, Seiberg:1994pq}. In the IR of softly broken $SU(N_c)$ gauged SQCD with $N_c$ colors and $N_f$ flavors, studied by some of the authors previously \cite{GuideToASQCD,Kondo:2021osz}, there exists a stable chiral symmetry breaking minimum for all $N_f<3N_c$, except for the specific case of $N_f=N_c$ in which the stability of the chiral symmetry breaking point cannot be determined. Such chiral symmetry breaking minima are expected to continuously connect to the true vacua of non-SUSY QCD as SUSY-breaking is increased.
Meanwhile, at the upper end of the free magnetic phase ($ \frac{3}{2} N_c\geq N_f\gtrsim 1.43N_c$) and into the conformal window, the theory suffers from an apparent runaway behavior, attributed to the tachyonic 2-loop masses of dual squarks. 
From the point of view of the electric description, we expect these runaways to end at an incalculable global minimum at large field values of $\mathcal{O}(\Lambda)$, while still leaving the chiral symmetry breaking minimum to be locally stable.

In this letter, we study QCD-like $Sp(N_c)$ theories, which are free of any baryonic runaways because there are no baryons. The superpotentials, moduli spaces and quantum vacuum structure of $\mathcal{N} = 1$ SUSY $Sp(N_c)$ theories have been worked out in \cite{IntriligatorSPn} by Intriligator and Pouliot, which will be the basis of our analysis for the AMSB perturbations \cite{Randall:1998uk, Giudice:1998xp}. The unitary symplectic group\footnote{Note that we use notation such that $U(N)\subset Sp(N)\subset SU(2N)$.} $Sp(N_{c})$ is the subgroup of $SU(2N_{c})$ which leaves invariant an antisymmetric tensor $J^{cd}$, which we can take to be $J = \textbf{1}_{N_{c}} \otimes i\sigma_{2}$. The dimension of this group is $N_{c}(2N_{c} + 1)$. We take for our matter content $2N_{f}$ fields $Q_{i}$, $i = 1\cdots 2N_{f}$, in the fundamental $2N_{c}$ dimensional representation of $Sp(N_{c})$. Similar to the conclusion of $SU(N_c)$ gauged theory, we predict that non-supersymmetric $Sp(N_c)$ QCD leads to chiral symmetry breaking all the way up to $N_f < 3(N_{c}+1)$. Our analysis cannot be applied to the range $N_f \geq 3(N_{c} +1)$ because the AMSB effects make the squark masses negative and thus the perturbed theory has no ground state. However, in the $Sp(N_c)$ theories, 2-loop level AMSB runaway in the dual squark direction is naturally lifted by the tree level SUSY potential at the upper end of the free magnetic phase ($\frac{3}{2} N_c\geq N_f\gtrsim 1.43N_c$) and the lower edge of the conformal window, and the $N_f=N_c+1$ meson point in the quantum-modified moduli space is also stable with respect to runaways. \\

Before we begin, we recall a well-known result from Anomaly Mediated Supersymmetry Breaking (AMSB), namely the tree-level supersymmetry breaking terms in the Lagrangian,
\begin{align}
    V_{\rm tree} = \partial_{i}Wg^{ij^{*}}\partial_{j}^{*}W^{*}+ m^{*}m\left(\partial_{i}Kg^{ij^{*}}\partial_{j}^{*}K - K\right) \nonumber\\
    + m\left(\partial_{i}Wg^{ij^{*}}\partial_{j}^{*}K - 3W\right) + c.c. \label{eq:genL}
\end{align}
where $g^{ij}$ is the inverse of the K\"ahler metric $g_{ij}=\partial_{i}\partial_{j}^{*}K$. For simplicity, we will always take the single spurion $m$ to be real\footnote{This corresponds to a choice of $U(1)_R$ frame, and can be exchanged for a different frame if convenient.}. When the K\"ahler potential $K$ is canonical in the fields $\phi_{i}$, this reduces to 
\begin{align}
	V_{\rm tree} &= m \left( \phi_{i} \frac{\partial W}{\partial \phi_{i}} - 3 W \right)
	+ c.c.
	\label{eq:tree}
\end{align}
This is accompanied by loop-level supersymmetry breaking effects in tri-linear couplings, scalar masses, and gaugino masses. The above expressions remain true at all energy scales due to the property of ultraviolet insensitivity for AMSB. 

\section{ADS Superpotential  $N_{f} \leq N_{c}$}
The space of vacua can be given a gauge invariant description in terms of arbitrary antisymmetric expectation values of the “meson” superfield $M_{ij} = Q_{ic}Q_{jd}J^{cd}$. In the low
energy theory, the dynamically generated superpotential restricted by
holomorphy and symmetries is the Affleck–Dine–Seiberg (ADS) superpotential 
\begin{equation}
W = (N_{c} + 1 - N_{f}) \left ( \frac{2^{N_{c} -1} \Lambda_{N_{c}, N_{f}}^{3(N_{c} + 1) - N_{f}}}{\Pf (M)}\right )^{1/(N_{c} + 1 - N_{f})}
\end{equation}
where we have set the root-of-unity factor $\omega_{N_{c} + 1 - N_{f}}$ to $1$ because it is not significant to our current discussion. To be concise, we will drop the subscripts on $\Lambda$ for the rest of this section. Subject to “D-flatness” constraints, the classical lagrangian has a moduli space of degenerate SUSY vacua $Q_{i}$ which when assembled in a single ($2N_f\times 2N_c$) matrix are of the form (up to gauge and global flavor rotations)
\begin{equation}
Q = 
\begin{pmatrix}
 a_{1} &  &  &  &  &  &\\
  & a_{2} &  &  &  &  &\\
  &  & \ddots &  &  &  &\\
  &  &  & a_{N_{f}} &  &  &
\end{pmatrix} \otimes \textbf{1}_{2}
\end{equation}
The corresponding matrix of Meson VEVs is
\begin{equation}
    M = \begin{pmatrix}
 0 & a_{1}^{2} &  &  & \\
 -a_{1}^{2} & 0 &  &  & \\
  &  & \ddots &  & \\
  &  &  & 0 & a_{N_{f}}^{2} \\
  &  &  & -a_{N_{f}}^{2} & 0 
\end{pmatrix}
\end{equation}
with $\Pf(M) = \prod_{i = 1}^{N_{f}} a_{i}^2$. In the region far away from the origin of moduli space (i.e. the meson VEVs are much larger than $\Lambda^{2}$), the theory is weakly
coupled and the Kähler potential is canonical in the quark superfields, so we can use \cref{eq:tree} to obtain 
\begin{multline}
    V = 
2\left| \frac{2^{N_{c}-1}\Lambda^{3N_{c} + 3 - N_{f}}}{\Pf(M)}\right|^{\frac{2}{N_{c} + 1 - N_{f}}} 
\left(\sum_{i=1}^{N_{f}}\left |\frac{1}{a_{i}}\right |^{2}\right)
\\ - m(3N_{c} + 3 - N_{f})\left(\frac{2^{N_{c}-1}\Lambda^{3N_{c} + 3 - N_{f}}}{\Pf (M)}\right)^{1/(N_{c}+1-N_{f})} + c.c.
\end{multline}
We use the inequality of arithmetic and geometric means to deduce that this potential is minimized when all the $|a_{i}|^{2}$ are equal. Thus, the global minimum can be found in the homogeneous direction   $M = a^{2} \delta_{ij} \otimes i\sigma_{2}$. Without AMSB, the vacuum clearly runs away to infinity. With AMSB, we can see that the runaway is lifted and we have a global minimum at 
\begin{equation}
a^{2} =
2^{\frac{N_c-1}{N_c+1}}
\Lambda^2
\left(
\frac{N_c+1}{3N_c+3-N_f}  \frac{\Lambda}{m}
\right) ^{ \frac{N_c+1-N_f}{N_c+1} }
\end{equation}
The minimum is indeed at $a^{2} \gg \Lambda^{2}$ for $m\ll\Lambda$, which justifies the weakly-coupled analysis. The $SU(2N_f)$ flavor symmetry is dynamically broken to $Sp(N_f)$, and whenever this happens we shall call the minimum a \textit{fully} chiral symmetry breaking minimum \footnote{Whenever the flavor symmetry is broken to a larger subgroup of $SU(2N_f)$, such as $Sp(\widetilde{N_c}) \otimes Sp(N_f - \widetilde{N_c} -2)$ in the dual squark branch (see later), we shall call the corresponding minimum a \textit{partial} chiral symmetry breaking minimum.}. The massless particle spectrum consists of the Nambu–Goldstone bosons, whose scalar and fermion
partners have masses that grow with $m$. Naively increasing $m$ beyond
$\Lambda$, one would expect that the only remaining degrees of freedom will be the massless NGBs.

\section{Quantum Modified Constraint  $N_{f} = N_{c} + 1$}
The dynamical superpotential is zero in this case. The low-energy degrees of freedom are the previously defined meson fields $M_{ij}$ subject to the quantum modified constraint
\begin{equation}
    \Pf(M) = 2^{N_{c}-1}\Lambda_{N_{c}, N_{c} + 1}^{2(N_{c}+1)} .\label{eq:QMC}
\end{equation}
There are no “baryons” in $Sp(N_c)$ because the invariant tensor $\epsilon^{c_{1}\cdots c_{N_{c}}}$ breaks up into sums of products of the $J^{cd}$, i.e. baryons break up into mesons. The quantum moduli space of vacua is smooth. The above constraint enforces that the meson fields have VEVs of $O(\Lambda)$. Therefore, higher order terms in the K\"ahler potential are not suppressed relative to the canonical term and the formula \eqref{eq:tree} cannot be trusted. So instead of implementing the quantum-modified constraint
in the superpotential via a Lagrange multiplier field $X$, we should perform a non-linear analysis using the constraint. For simplicity, we will use units where $\Lambda = 1$. In general the equivalence classes of the $2N_f\times 2N_f$ anti-symmetric meson field $M$ under $SU(2N_f)$ flavor symmetry can be represented by 
\begin{align}
    M = \text{diag}\{ a_1, \cdots ,a_{N_f}\}\otimes i\sigma_2,
    \label{eq:Meson flavor class}
\end{align}
in which $a_1$ to $a_{N_f}$ are non-negative real numbers that satisfy quantum modified constraint $\prod_{i=1}^{N_f}a_i = 2^{N_c-1}$.

The moduli space contains a special point of enhanced symmetry: the meson point with  $M_{ij} = \alpha \delta_{ij}\otimes i\sigma_{2}$, where $\alpha = 2^{(N_{c}-1)/(N_{c}+1)}$. We shall first perform AMSB around this point.

\subsection{The Meson Point}

To satisfy the constraint \eqref{eq:QMC} at the meson point and examine stability only along the direction that $M$ stays in the form \eqref{eq:Meson flavor class}, we make
the change of variables
\begin{equation}
    M = \alpha e^{\Pi} \otimes i\sigma_{2} = \alpha(1 + \Pi + \frac{1}{2} \Pi^{2} \cdots) \otimes i\sigma_{2}
\end{equation}

where $\Pi$ is a diagonal, traceless matrix, so that $\det(e^{\Pi}) = 1$.
The pure imaginary part of this $\Pi$ will turn out to be Goldstone modes. 
The K\"ahler potential is built out of flavor invariants, e.g. Tr($M^{\dagger}M$), (Tr ($M^{\dagger}M$))$^{2}$, Tr($M^{\dagger}MM^{\dagger}M$), etc. Notice that they will all contribute at quadratic order in the hadron superfields and will have similar algebraic form. For example, let’s examine the $\Pi$ contribution of the aforementioned three terms up to quadratic order (an overall factor of 2 appears in the traces due to the tensor product with $i\sigma_{2}$):
\begin{align}
    \Tr M^{\dagger}M = 2(\Tr \Pi^{\dagger}\Pi + \frac{1}{2}\Tr \Pi^{2} + \frac{1}{2}\Tr \Pi^{\dagger 2}) \label{eq:MM} \\
    (\Tr M^{\dagger}M)^{2} = 8(\Tr \Pi^{\dagger}\Pi + \frac{1}{2}\Tr \Pi^{2} + \frac{1}{2}\Tr \Pi^{\dagger 2}) \\
    \Tr M^{\dagger}MM^{\dagger}M = 8(\Tr\Pi^{\dagger}\Pi + \frac{1}{2}\Tr \Pi^{2} + \frac{1}{2}\Tr \Pi^{\dagger 2}) \label{eq:MMMM}
\end{align}

A useful formula here will be the tree level AMSB potential corresponding to $K = \varphi^{\dagger}\varphi + \eta /2(\varphi^{2} + \varphi^{\dagger 2}$). Using the general formula \eqref{eq:genL}, we then get
\begin{align}
    V_{\text{AMSB}} &= \eta^{2}m^{2}\varphi^{\dagger}\varphi + \frac{\eta}{2}m^{2}(\varphi^{2} + \varphi^{\dagger 2}) \nonumber \\
    &= (\eta^{2} + \eta)m^{2}(\Re (\varphi))^{2} 
    + (\eta^{2} - \eta)m^{2}(\Im (\varphi))^{2} \nonumber
\end{align}
Setting $\eta = 1$ corresponds to the K\"ahler potential (upto factors of 2) for each component of $\Pi$ in \eqref{eq:MM}-\eqref{eq:MMMM}, so that the $\Im(\Pi)$ are the massless pions --- the Goldstone bosons of broken chiral flavor symmetry. 

For $\Im(\Pi)$ to remain massless, Goldstone’s theorem ensures that all
higher-order meson flavor invariants of the K\"ahler potential will give
contributions proportional to the right-hand-sides of \eqref{eq:MM}-\eqref{eq:MMMM}. Moreover, they will (in aggregate) come with a positive
sign in order for the $\Pi$ to have a physical kinetic term (i.e. a positive-definite K\"ahler metric).
Also note that this will result in a positive mass for $\Re (\Pi)$, which will stabilize this direction. Thus, unlike the AMSB-deformed $SU(N_{c})$ SQCD theory where baryonic contributions lead to incalculable minima, the meson point here is stable. \\

\subsection{Away from the Meson Point}

Next, we consider the possibility of having another stable minimum away from the meson point. First we examine this possibility with just a canonical K\"ahler potential. Then we investigate whether higher order K\"ahler terms could change the conclusion.

Restricting to a canonical K\"ahler potential, we show that any point in the quantum modified moduli space other than the meson point would not be a stable minimum, thus suggesting that the meson point is a global and chiral symmetry breaking minimum. To see this instability, we perturb around an arbitrary point 

\begin{equation}
    M = A e^{\Pi} \otimes i\sigma_{2} = A(1 + \Pi + \frac{1}{2} \Pi^{2} \cdots) \otimes i\sigma_{2}
\end{equation}
in the moduli space, in which we have taken $A=$ diag$(a_1,a_2,\cdots,a_{N_f})$ to have different eigenvalues satisfying $\prod_{i=1}^{N_f}a_i=\alpha^{N_f}$, and $\Pi$ again to be a traceless matrix.
We expand $\Pi=\Pi^A T^A$ with hermitian $SU(N_f)$ generators $T^A$ and complex scalar fields $\Pi^A$. Focusing on contributions from $\Tr M^\dagger M$, we find a $\Pi$-linear term $K\supset \Tr M^\dagger M \supset 2(\Tr A A^\dagger T^A)(\Pi^A+{\Pi^A}^\dagger)$
in the K\"ahler potential. This $\Pi$-linear term becomes zero only when the matrix $A$ has identical eigenvalues.
This linear term in $K$ leads to also a $\Pi$-linear term in the scalar potential
\begin{align}
    V_{AMSB}\supset \Pi^E (-\frac{1}{2}v^E
    +\frac{1}{2}\mathcal{C}^{AE}\mathcal{C}^{-1}_{AB}v^B
    +v^A \mathcal{C}^{-1}_{AB}\mathcal{C}^{EB}\nonumber\\
    -\frac{1}{4}v^A \mathcal{C}^{-1}_{AC} (\mathcal{D}^{CDE}+\mathcal{D}^{ECD})\mathcal{C}^{-1}_{DB}v^B)
\end{align}
where $v^A=2\Tr (AA^\dagger T^A)$, $\mathcal{C}^{AB}=\Tr (AA^\dagger T^A T^B)$, and $\mathcal{D}^{ABC}=\Tr (AA^\dagger T^A T^B T^C
)$. Also note that there is a $\Pi^\dagger$ linear term of a similar form in  $V_{AMSB}$.

In general, it is hard to find a necessary and sufficient condition for this linear term to be vanishing. But it's straightforward to see that it disappears when $v^A$, or equivalently, when the $\Pi$-linear term in $K$ is zero. 
There is also a constant term in $V_{AMSB}$, which is 
\begin{equation}
    V_{AMSB} \supset \frac{1}{2}v^A \mathcal{C}^{-1}_{AB}v^B-2\Tr(AA^\dagger)
\end{equation}
This term is most negative when at the meson point. So overall, the nonzero linear term in $V_{AMSB}$, showing up everywhere in the moduli space other than the meson point, ensures that we have a unique, stable, fully chiral symmetry breaking minimum at the meson point.

\begin{figure}
    \centering
    \includegraphics[width=0.75\linewidth]{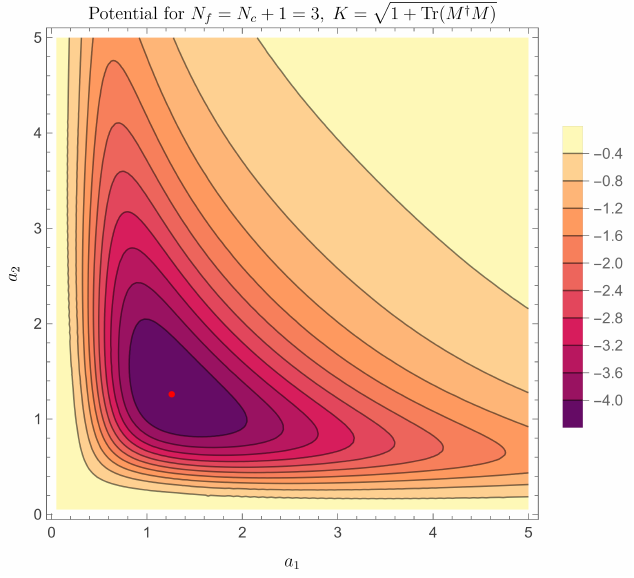}
    \caption{The potential (in $\Lambda=1$ units) for the case $N_f=N_c+1=3$, with K\"ahler potential $K=\sqrt{1+\Tr{M^\dagger M}}$ which properly interpolates between the small- and large-field regimes and maintains positive-definite kinetic terms everywhere. The red dot indicates the meson point, which can be seen to remain the global minimum of the theory. We parametrize the moduli as in \cref{eq:Meson flavor class} with $a_3 = \frac{2}{a_1a_2}$ to satisfy the quantum modified constraint.}
    \label{fig:sqrtK}
\end{figure}

Next we consider whether the inevitable higher order terms in the K\"ahler potential could change this conclusion. We require that the second derivative of the K\"ahler potential $\partial^2 K/ \partial \phi^i \partial \phi^{*j}$ be positive definite on the \textit{entire} moduli space for ensuring a positive definite kinetic term for the mesons. This is rather constraining, such that generically adding higher-order K\"ahler terms truncated at some finite order results in a non-positive-definite K\"ahler metric. We expect the \textit{full} K\"ahler potential to interpolate between the canonical form for the mesons $\Tr{M^\dagger M}$ at small field values and that for the quarks $\sqrt{\Tr{M^\dagger M}}$ at large field values (where the perturbative UV description should apply). So as an ansatz we take $K=\sqrt{1+\Tr{M^\dagger M}}$. Numerical computations at low values of $N_f$ show that this ansatz satisfies the requirement of positive-definiteness, and the meson point remains the global minimum (see Figure \ref{fig:sqrtK} for the $N_f=3$ example). Any deformations to the canonical K\"ahler potential we have tried that produce other minima or runaways suffer a lack of positive-definiteness of the kinetic term, and thus are not physically acceptable. 

Taken all together, we have strong evidence supporting the conjecture that the meson point is the true global minimum of the full theory.

\section{S-Confinement $N_{f} = N_{c} + 2$}
 The case $N_c=1$ is equivalent to $SU(2)$ s-confining SQCD with AMSB, which in \cite{delima2023sconfining} was found to not have a chiral symmetry breaking minimum for small field values. In \cite{Bai:2021tgl}, a three-generation SUSY $SU(5)$ GUT with AMSB also had a classically conformal superpotential and was also found to not have a symmetry breaking minimum. Whether an additional deformation of such theories could salvage the situation remains an open question, but we do not investigate it further in this work. \\

For the case of $N_{c} > 1$, we find a stable chiral symmetry breaking
minimum and no runaway directions. At leading order we take a canonical K\"ahler potential for the low energy meson field $M$, which is justified when the eigenvalues of $M$ are much smaller than the dynamical scale $\Lambda$ (i.e. the theory is strongly coupled). The superpotential is
\begin{equation}
    W = - \kappa\ \Pf(M),
\end{equation}
where we use units with $\Lambda=1$ and 
$\kappa$ is an unknown order 1 coupling that makes the K\"ahler potential canonical.
For $M$, we start with a block diagonal matrix with arbitrary (possibly different) entries $v_{i}$ for each block. For fixed $\Pf(M)$, any terms outside the $2 \times 2$ blocks would increase $V_{SUSY}$, justifying their omission. Also, given that we are taking $m$
real, it is enough to look for minima with all fields $v_{i}$ real. Later we deduce that the global minimum occurs when the VEVs are homogeneous, namely $M_{ij} = v \delta_{ij} \otimes i\sigma_{2}$. Using equation \eqref{eq:tree}, and that in general $\Pf(M) = \prod_{i = 1}^{N_{f}}v_{i}$, we get 
\begin{align}
    V =& \frac{1}{2}\left(\kappa \Pf(M)\right)^{2}\left(\sum_{i=1}^{N_{f}}\frac{1}{|v_{i}|^{2}}\right)
    + 2m\kappa(3 - N_{f}) \Pf(M).
\end{align}

For a fixed $\Pf(M)$, the inequality of arithmetic and geometric means tells us that this function is minimized when all the $v_{i}$ are equal. Setting all the $v_{i}$ to $v$ in the above potential and minimizing, we get
\begin{equation}
    v = \left(\frac{2m(N_{c} - 1)}{\kappa(N_{c}+1)}\right)^{1/N_{c}}.
\end{equation}
The potential at this minimum is a long expression, but in particular it is 
\begin{equation}
    V_{min} = -\mathcal{O}(m^{(2N_{c}+2)/N_{c}})
\end{equation}
This is the fully chiral symmetry breaking minimum that we hope to be continuously connected to that of non-SUSY $Sp(N_{c})$ QCD. Note that since there is no Yukawa-like term in the superpotential (unlike the corresponding $SU(N_{c})$ case), we don't get a 2-loop mass for the meson. So there is no 2-loop potential that could destabilize the chiral symmetry breaking minimum.

We should finally check the effects of higher order terms in the K\"ahler potential, the leading ones being (Tr $M^{\dagger}M$)$^{2}$ and Tr $M^{\dagger}MM^{\dagger}M$ with unknown coefficients (including signs). Using \eqref{eq:genL}, we find that these give potential terms $\sim m^{2}v^{4}$. At the value of $v \sim m^{1/N_{c}}$ that minimizes the potential, these are higher order in $m$ (as $2 + 4/N_{c} > 2 + 2/N_{c}$) and can be neglected.

\section{Free Magnetic Phase $N_{c} + 2 \leq N_{f} \leq \frac{3}{2} (N_{c}+1)$}

For this range of flavors, the SUSY theory is in the free magnetic phase and the IR is described by an $Sp(N_f-N_{c}-2 \equiv \widetilde{N}_c)$ gauge theory with $2N_f$  dual quarks $q^i(i=1\dots 2N_f)$ in the fundamental representation (i.e. $N_f$ ``flavors") and a  gauge-singlet antisymmetric meson $M_{ij}$ ($=-M_{ji}$). The theory has a global $SU(2N_f)$ flavor symmetry with $q$ in the $(\overline{2N_f})$ and $M$ in the $(N_f(2N_f-1))$ representations. The superpotential of this free magnetic theory is given by
\begin{align}
	W = \frac{\lambda}{2} M_{ij} q^{i}_{c} q^{j}_{d}J^{cd},\label{eq:W_yukawa}
\end{align}
in which all the fields are already normalized to have  K\"ahler potential $\frac{1}{2}\Tr M^{\dagger}M + \Tr q^{\dagger}q$. Also note that only the deep
IR behavior of the theory is specified and we do not have
control over the relative strengths of the gauge interaction
and the Yukawa interaction $\lambda$ in \eqref{eq:W_yukawa}. \footnote{It should be noted that at exactly $N_f = \frac{3}{2}N_c$, the theory enjoys classical conformal invariance. This is known to obstruct the usefulness of AMSB \cite{Bai:2021tgl, delima2023sconfining}, and so this particular point remains an open problem. One may, however, speculate about chiral symmetry breaking in the non-SUSY theory based on the results immediately above and slightly below in $N_f$.} \\

One of the most significant properties of $Sp(\widetilde{N}_{c})$ gauge theory, different from $SU(\widetilde{N}_{c})$, is that for the entire free magnetic phase window the baryonic runaways of the latter are absent in the former, regardless of the negative 2-loop mass of quarks when $N_f \gtrsim 1.43 N_{c}$. Furthermore, throughout the free magnetic window, the chiral symmetry breaking minimum is stable and likely to be the global minimum of the theory. To reach this conclusion, we have to carefully check several branches presented below case by case.

We proceed by first analyzing the dual quark direction, where the entire dual gauge group is Higgsed by giving VEVs to the dual squarks. As mentioned, the theory is free of runaways in this direction throughout the whole free magnetic phase.
We next exhibit the chiral symmetry breaking minimum along the mesonic direction.
Finally, we check the mixed directions, where only some meson VEVs are turned on, to ensure that they contain no runaways.

\subsection{RG Analysis}

We first present some general renormalization results from \cite{MartinVaughn, deGouvea:1998ft} when the superpotential is \eqref{eq:W_yukawa}, which will be useful in both the free magnetic phase and conformal window sections.

For a more general Yukawa-like superpotential $W = \frac{1}{3!} \lambda_{ijk} \phi_i \phi_j \phi_k$, the anomalous dimension for scalar fields are 
\begin{align}
	\gamma_{i} = \mu\frac{d}{d\mu} \ln Z_{i}(\mu)
	= \frac{1}{8\pi^2} (2g^2 C_i - \frac{1}{2} \sum_{j,k}\lambda_{ijk}^*\lambda_{ijk}).
\end{align}

Substituting in our specific Yukawa potential \eqref{eq:W_yukawa}, we get the 1-loop anomalous dimensions
\begin{align}
	\gamma_q &= \frac{1}{8\pi^2} (2 C_F g^2 - (2 N_f-1)\lambda^2) \label{eq:anomalous_q} \\
	\gamma_M &= -\frac{2 \widetilde{N}_c \lambda^2}{8\pi^2} \label{eq:anomalous_M}
\end{align}

where $C_F = (2\widetilde{N}_c + 1)/4 $ is the quadratic Casimir of the dual gauge group $Sp(\widetilde{N}_{c})$ in its fundamental representation.

For the beta functions of the gauge coupling and the Yukawa coupling, we have their exact results and 1-loop approximations as
\begin{align}
    \beta_g =-\frac{g^4}{8\pi^2} \frac{3(\widetilde{N}_{c}+1)-N_f-N_f \gamma_q}{1-(\widetilde{N}_{c}+1)g^2/(8\pi^2)}  
    \overset{\mathrm{1-loop}}{=}\frac{(-\widetilde{b})g^4}{8\pi^2}  \label{eq:beta_g^2_exact}
\end{align}
\begin{align}
    \beta_\lambda =&-(\gamma_M + 2\gamma_q) \lambda^2 
    \\
\overset{\rm{1-loop}}{=}& \frac{\lambda^2}{8\pi^2}((2\widetilde{N}_{c}+4N_f-2)\lambda^2-(2\widetilde{N}_{c}+1)g^2) \label{eq:beta_lambda^2_exact}
\end{align}

where $\widetilde{b}=3\widetilde{N}_{c}+3-N_f$.
Note that we follow the notation $\beta_g= \mu \frac{d}{d\mu} g^{2}$ etc. \\

In a small neighborhood of the origin of moduli space, the theory is allowed to run into the deep IR. Then since the magnetic theory is IR free, both gauge coupling $g$ and Yukawa coupling run to zero. At 1-loop,
the coupled beta functions have the relation
\begin{align}
    \frac{\beta_g}{g^2}-\frac{\beta_\lambda}{\lambda^2} =& \frac{N_f-\tilde{N}_c-2}{8\pi^2}g^2 - \frac{\tilde{N}_c+2N_f-1}{4\pi^2}\lambda^2
    \label{ratio_run}
\end{align}

We can solve the RGE, and we find that in the deep IR
\begin{align}
    \frac{g^2}{8\pi^2} \sim& \frac{1}{(3\tilde{N}_c+3-N_f)\ln \mu} \\
    \frac{\lambda^2}{4\pi^2} \sim& \frac{N_f-\tilde{N}_c-2}{(3\tilde{N}_c+3-N_f)(\tilde{N}_c+2N_f-1) \ln\mu}
\end{align}

Plugging these into \eqref{ratio_run}, we can conclude that close to the trivial fixed point the couplings approach along the trajectory given by  
\begin{equation}
    \frac{d}{d \log \mu}\frac{g^{2}}{\lambda^{2}} = 0 \implies \frac{\beta_{g}}{g^{2}} = \frac{\beta_{\lambda}}{\lambda^{2}}.
\label{eq:coupled_beta}
\end{equation}
This relation allows $\lambda$ to be written in terms of $g$ as
\begin{align}
    \frac{\lambda^2}{g^2}=\frac{N_f-\widetilde{N}_{c}-2}{2\widetilde{N}_{c}+4N_f-2}
\end{align}

when close to the IR fixed point.

\begin{figure}[t]
\includegraphics[width=0.82\columnwidth]{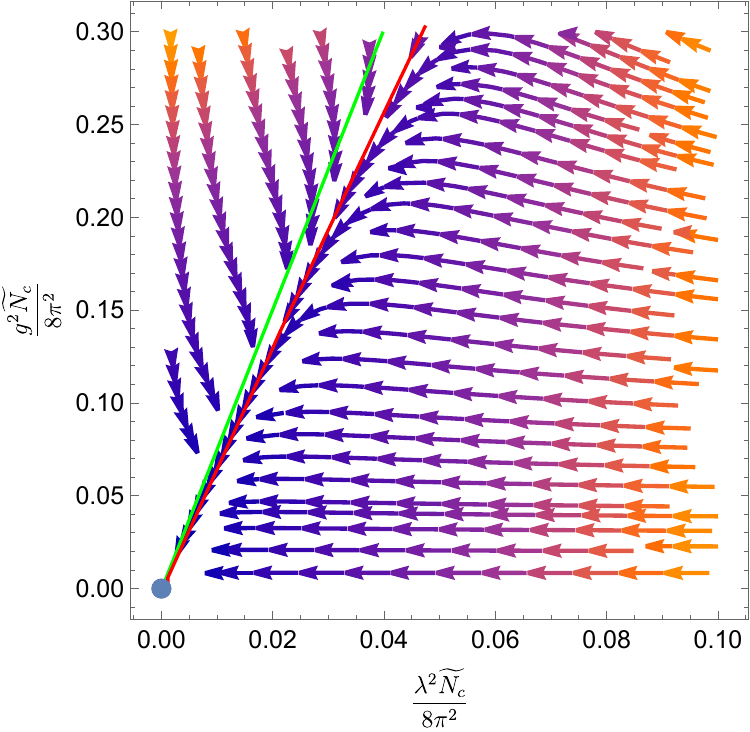}\\ 
\caption{A plot for $N_f = 330$, $N_c = 230$, i.e. $N_f = 1.429(N_c+1)$. Couplings clearly flow to the fixed point (0, 0) along a straight-line trajectory. The red line is a plot of $m_{q}^{2} = 0$ and indeed the trajectory converges onto it near (0, 0). The region below the green line satisfies $m_{M}^{2} > 0$; the trajectory completely lies in this region.}
\label{fig:FreeRG}
\end{figure}

Now, through anomaly mediation, scalars acquire 2-loop masses given by
\begin{align}
	m_{i}^{2}(\mu) &= - \frac{1}{4} \dot{\gamma}_{i}(\mu) m^{2} 
	\label{eq:m2_anomaly_scalar} 
\end{align}
in which $\dot{\gamma} = \mu \frac{d}{d\mu} \gamma_{i}$ (in general, physical masses are the sum of contributions from the tree-level or non-perturbative superpotential, tree-level AMSB and loop-level AMSB). The signs of these mass formulae have been chosen consistently with the sign convention of the $\gamma$'s.
Thus, the 2-loop masses of the dual quarks and the mesons are \cite{Kondo:2021osz}
\begin{align}
	m_q^2 & = \frac{(-\widetilde{b})g^4}{(16\pi^2)^2} \frac{2N_f^2-(6\widetilde{N}_{c}+7)N_f-2\widetilde{N}_{c}^2+2\widetilde{N}_{c}+3}{2(\widetilde{N}_{c}+2N_f-1)} m^2
	, \label{eq:2loopmass_q}\\
	m_M^2 & = \frac{(-\widetilde{b}) 2\widetilde{N}_{c} g^2 \lambda^2 }{(16\pi^2)^2} m^2 \label{eq:2loopmass_M}
\end{align}

where $\widetilde{b}$ is negative (in the free magnetic phase), ensuring that the meson masses remain positive throughout the free magnetic window. The dual quarks have positive masses for most of the window, however become negative at the upper end when $N_f\gtrsim 1.43 (N_{c}+1)$ 
(for large $N_{c}$). The RG flow is illustrated in Figure \ref{fig:FreeRG} for a choice of $N_{f} \approx 1.43(N_{c} +1)$, where we expect $m_{q}^{2}$ to switch sign (and hence be $\approx 0$) while $m_{M}^{2}$ remains positive.

\subsection{Dual Quark Branch}
For $N_f\gtrsim 1.43 (N_{c}+1)$, we give  D-flat VEVs to the dual squark. The squark VEV appears to trigger condensation of the mesons as well, leading to a minimum along a mixed branch. We focus on the slice of moduli space parametrized by the VEVs $q = \phi_q \textbf{1}\otimes \textbf{1}_2$ and $M = \phi_M \textbf{1}\otimes i\sigma_2$. \footnote{If we just used the available gauge/flavor rotations and D-flat conditions, generally there could be more non-zero entries in the $q$ matrix, whose presence in the scalar potential could be parametrized using a CKM-esque unitary matrix. However, even a numerical exploration turns out to be very challenging if we assume full generality. Therefore, we resort to a numerical exploration on a slice with enhanced symmetry.} The superpotential and K\"ahler potential are then
 
 \begin{align}
	W = \frac{\lambda}{2} M_{ij} q^{i}_{c} q^{j}_{d}J^{cd} = \lambda \widetilde{N}_{c}\phi_M \phi_q^{2}\label{eq:W_yukawa}
\end{align}
\begin{align}
	K = \frac{1}{2}\Tr M^{\dagger}M + \Tr q^{\dagger}q = N_f \left|\phi_M \right|^{2} + 2 \widetilde{N}_{c}\left|\phi_q \right|^{2}
\end{align}
which lead to the scalar potential 
\begin{align}
  V&=  -N_fm^2\frac{\dot{\gamma}_M}{4} |\phi_M|^2
+\frac{1}{N_f} \left| \frac{\partial W}{\partial\phi_M} \right|^2
- 2\widetilde{N}_{c}m^2\frac{\dot{\gamma}_q}{4} |\phi_q|^2 + \frac{1}{2\widetilde{N}_{c}} \left| \frac{\partial W}{\partial\phi_q} \right|^2 
\nonumber\\
&+3m(W+h.c.) -m \left( 1-\frac{\gamma_M}{2} \right)\left(\phi_M\frac{\partial W }{\partial \phi_M }+h.c.\right) -m \left( 1-\frac{\gamma_q}{2} \right)\left(\phi_q\frac{\partial W }{\partial \phi_q }+h.c.\right)  \nonumber\\
&=  -N_fm^2\frac{\dot{\gamma}_M}{4} \phi_M^2
+\frac{(\lambda\widetilde{N}_{c} \phi_q^{2})^2}{N_f}
- 2\widetilde{N}_{c}m^2\frac{\dot{\gamma}_q}{4} \phi_q^2
+\frac{(2\lambda\widetilde{N}_{c}\phi_M \phi_q)^2}{2\widetilde{N}_{c}} +m \left(\gamma_M + 2\gamma_q\right)\left(\lambda\widetilde{N}_{c}\phi_M \phi_q^{2}\right)
\end{align}

\begin{figure}[t]
\centering
\includegraphics[width=0.75\columnwidth]{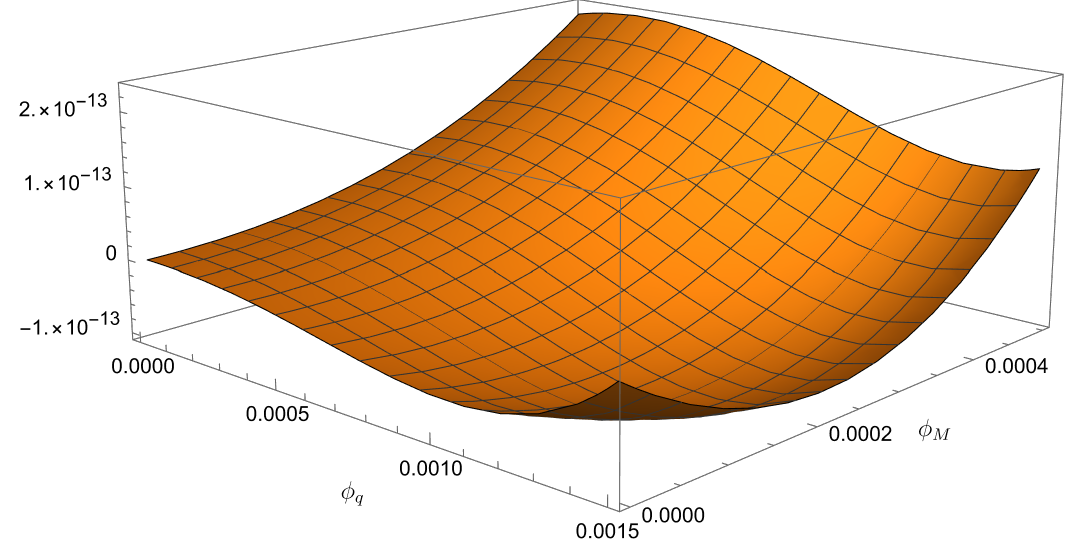}\\ 
\includegraphics[width=0.75\columnwidth]{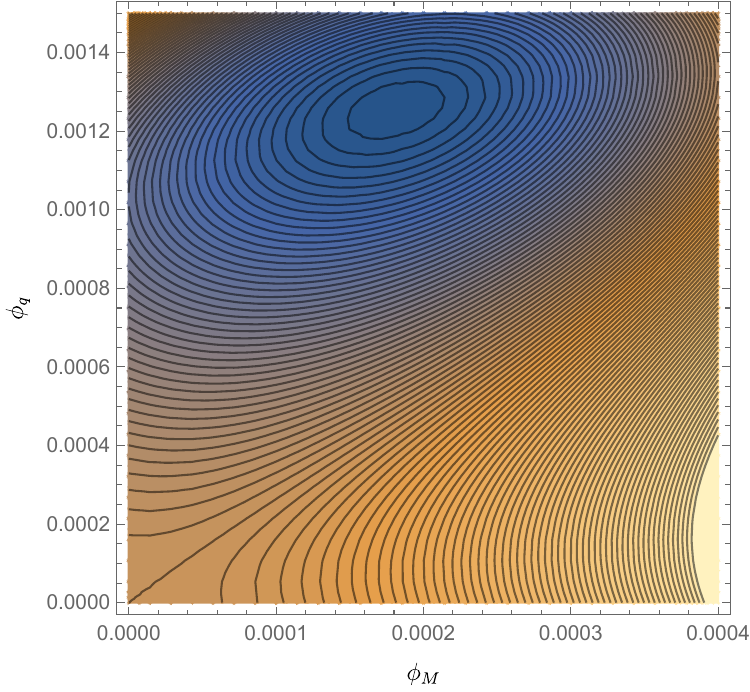}\\
\caption{A 3D plot and a contour plot of the scalar potential in the upper end of the free magnetic phase $1.43(N_c + 1) < N_f < 1.5(N_c + 1)$ when the dual squarks are tachyonic and the meson mass-squared is positive. Both the dual squark VEV $\phi_q$ and the meson VEV $\phi_M$ are non-zero at the unique minimum. For these plots, we have assigned the numerical values $\widetilde{N_c} = 6$, $N_f = 22$ and $\frac{m}{\widetilde{\Lambda}} = 0.1$. The value of the scalar potential at the minimum is $-1.0305 \times 10^{-13}$ in units where $\widetilde{\Lambda} = 1$.}
\label{fig:FreeContour}
\end{figure}

Note that even though the terms quadratic and linear in $\phi_M$ may be positive, it is not necessary that the minimum of the potential would be for $\phi_M = 0$. So $\phi_M$ should not be integrated out {\it a priori}\/. In the deep IR, we can use equations (5.3), (5.4), (5.9), (5.10), (5.14) and (5.15) to express the scalar potential as a function of $N_f$, $\widetilde{N_c}$, $m$, $\phi_M$ and $\phi_q$ only. Even for the slice that we have chosen, the scalar potential is complicated and an analytical derivation of the minimum does not seem to be possible. The scalar potential is numerically analyzed using contour plots on the $\phi_M-\phi_q$ plane, for different values of the ``unconstrained" parameters $\widetilde{N_c}$, $N_f$ and $\frac{m}{\Lambda}$. We take the RGE scale $\mu \approx \frac{1}{2}(\phi_M + \phi_q)$ (one could also take the geometric mean without changing the net qualitative outcome of the analysis). Regardless of these choices, we always find a single minimum at non-zero $\phi_M$ and $\phi_q$ (hence the ``mixed branch"), where the original $SU(2N_f)$ chiral flavor symmetry would be broken into $Sp(\widetilde{N_c}) \otimes Sp(N_f - \widetilde{N_c} -2)$. In particular, there is no runaway despite tachyonic dual squarks, and also no minimum is seen for this potential along the pure dual squark or pure mesonic branch (i.e. for either $\phi_M = 0$ or $\phi_q = 0$). For illustration, a 3D plot and contour plot have been shown in Figure \ref{fig:FreeContour} along with the corresponding numerical choices made. For each set of parameters, we locate the coordinates of the minimum in moduli space and evaluate the value of the scalar potential at the minimum. However, we are unable to provide an analytical expression. \\

Interestingly, we will show in the next subsection that the \textit{partial} chiral symmetry breaking minimum from this branch is only a local miminum 
for $1.43(N_c + 1) < N_f < 1.5(N_c + 1)$, with the global minimum coming from a pure mesonic branch after accounting for non-perturbative dynamics. 

\subsection{Mesonic Branch}

For $N_c + 2 \leq N_f\lesssim 1.43 N_{c}$, where $m_{M}^{2}$ and $m_{q}^{2}$ are positive, all the scalar potential terms in (5.18) that are quartic and quadratic in $\phi_q$ would be positive (the sign of the term linear in $m$ could be flipped by doing a $U(1)_R$ rotation on $m$), and there is no term linear in $\phi_q$, therefore allowing us to integrate out the dual squarks by taking $\phi_q = 0$. Without their effects, the beta function of the gauge theory flips sign, allowing the theory to generate a new IR dynamical scale given by
\begin{equation}
    \Lambda_L^{3(\widetilde{N}_{c}+1)} = 2^{N_f} \widetilde{\Lambda}^{3(\widetilde{N}_{c}+1) - N_f} \Pf(\lambda M).
\end{equation}

Thus the usual superpotential of pure SYM is generated as
\begin{align}\begin{split}
    W &= (\widetilde{N}_{c}+1) 2\cdot 2^{\frac{-2}{\widetilde{N}_c+1}}\Lambda_L^3\\
&=-(\widetilde{N}_{c}+1)
    (2^{N_f + \widetilde{N_c}-1} \Pf(\lambda M))^{\frac{1}{\widetilde{N}_{c}+1}}
\end{split}\end{align}
Note that we set $\widetilde{\Lambda}=\omega_{2\frac{\widetilde{N}_{c}+1}{3(\widetilde{N}_{c}+1)-N_f}}$, namely only keeping its important phase factor from scale matching. 

We give the meson a VEV with full rank. Upon adding tree level AMSB term, and by doing an analysis similar to the ADS section ($N_f<N_{c}$),  one can find minimum along the homogeneous direction $M=v\textbf{1}\otimes i\sigma_2$ with the potential
 \begin{equation}
     V=N_f 2^{2\frac{N_f + \widetilde{N_c}-1}{\widetilde{N_c}+1}} |\lambda v|^{2(\frac{N_f}{\widetilde{N}_{c}+1}-1)}
     -2m(N_f-3(\widetilde{N}_{c}+1))
      2^{\frac{N_f +\widetilde{N_c}-1}{\widetilde{N_c}+1}} (\lambda v)^{\frac{N_f}{\widetilde{N}_{c}+1}} 
 \end{equation}
at the point
\begin{equation*}
    v=2^{-\frac{N_f+\widetilde{N}_c-1}{N_f-2(\widetilde{N}_c+1)}}\left(\frac{1}{\lambda}\right)
    \left( \frac{N_f- 3(\widetilde{N}_{c}+1) }{ N_f-(\widetilde{N}_{c}+1) }m\right)^{ \frac{\widetilde{N}_{c}+1}{N_f-2(\widetilde{N}_{c}+1)} },
\end{equation*}
\begin{align}
    V_{min} = -\mathcal{O} \left( m^{2\frac{N_f-(\widetilde{N}_{c}+1)}{N_f-2(\widetilde{N}_{c}+1)}} \right) .
\end{align}

At this fully chiral symmetry breaking minimum, the original $SU(2N_f)$ chiral flavor symmetry is completely broken into $Sp(N_f)$ by meson VEVs. Since the 2-loop mass of meson from \eqref{eq:2loopmass_M} goes as $m^2 v^2$, it also contributes to the potential at the same order in $m$, however it is loop-suppressed. Thus, we find that the minimum is stable. \\

Now for $1.43(N_c + 1) < N_f < 1.5(N_c + 1)$, we could carry out the same derivation to get a purely mesonic branch minimum, provided we can justify integrating out the dual squarks which have a negative 2-loop AMSB mass. Naively looking at equation (5.18), the term quadratic in $\phi_q$ could have either sign when $m_q^2 < 0$, and it is not conclusive as to whether $\phi_q = 0$ minimizes the scalar potential. But it should be noted that non-perturbative effects are not included in this potential. So we shall do {\it a posteriori}\/ justification as follows. We will integrate out the dual squarks, derive an expression for $\phi_M$ that minimizes the low energy scalar potential (5.21) (which includes non-perturbative contributions like gaugino condensation), plug that $\phi_M$ back into (5.18) to get an effective potential for $\phi_q$, and then show that it has a minimum at $\phi_q = 0$ (or in other words $q$ has a net positive mass-squared that allows integrating it out consistently). However, this process could only be carried out numerically rather than analytically because it involves the mixed branch potential again. \\

Continuing the example used in Figure \ref{fig:FreeContour}, let us discuss the case where $\widetilde{N_c} = 6$, $N_f = 22$ and $\frac{m}{\widetilde{\Lambda}} = 0.1$. We use equation (5.10) to substitute for $\lambda$ in the deep IR, and we take the RGE scale $\mu \approx \phi_M$. Then the minimum is found to be at $v = 0.00516$, in units where $\widetilde{\Lambda} = 1$, which is consistent with the units in the mixed branch analysis. So we plug in $\phi_M = 0.00516$ into equation (5.18), and use equations (5.3), (5.4), (5.9), (5.10), (5.14) and (5.15) to express this effective scalar potential as a function of $\phi_q$ only. We find that $\phi_q = 0$ is indeed a minimum and hence conclude that it was justified to integrate out the dual squarks in the first place. It remains to check whether the pure mesonic minimum is deeper than the mixed $q$ \& $M$ minimum, and we find that indeed it is. In this example, the pure mesonic potential has a minimum value of $-5.138\times 10^{-10}$, which is much deeper than the mixed brach potential's minimum value of $-1.0305 \times 10^{-13}$ (in the same units). \\

Based on many numerical explorations similar to the example presented above, going up to $\widetilde{N_c} = 15$, it seems very convincing that the pure mesonic branch provides the global (fully) chiral symmetry breaking minimum throughout the free magnetic phase. The mixed $q$ and $M$ branch features a local minimum when the dual squarks turn tachyonic.

\subsection{Mesonic Mixed Branches}

Instead of turning on all of the meson VEVs, we can choose to turn on only some of them. These will reveal tree-level AMSB contributions within the free magnetic
phase. We will ignore $\lambda$ in this subsection as it is nit important for the discussion here.

We begin by splitting meson matrix into 
\begin{align}
    M 
    &= \begin{pmatrix}
\widetilde{M}_{R_f\times R_f} &0\\
 0 & \widehat{M}_{(N_f-R_f)\times (N_f-R_f)} 
\end{pmatrix} \otimes i\sigma_2
\end{align}
and without loss of generality, we look for minima at diagonal
$\widetilde{M}$ and $\widehat{M}$. We then give the lower component $\widehat{M}$ a VEV.
This gives masses to $2(N_f-R_f)$ dual quarks, leaving the $Sp(\widetilde{N}_{c})$ gauge theory with $R_f$ massless flavors and a new dynamical scale
\begin{equation}
    \Lambda_L^{3(\widetilde{N}_{c}+1)-R_f} = 2^{N_f - Rf}\widetilde{\Lambda}^{3(\widetilde{N}_{c}+1) - N_f} \Pf(\widehat{M})
\end{equation}
with $\widetilde{\Lambda}$ being the Landau pole of the  $Sp(\widetilde{N}_{c})$ dual theory. To avoid congestion, in this subsection we set 
\begin{equation*}
    2^{N_f - Rf}\widetilde{\Lambda}^{3(\widetilde{N}_{c}+1) - N_f}=1
\end{equation*} 
Finally, we analyze the region where $\widetilde{M}$ is small compared to the $\widehat{M}$ VEVs as well as the generated scale $\Lambda_L$. \\

For $1\leq R_f< \widetilde{N}_{c}+1$, the remaining theory is of ADS-type and has the superpotential
\begin{align}
    W=(\widetilde{N}_{c}+1 - R_f)
    \left ( \frac{2^{\widetilde{N}_{c} -1} \Lambda_{L}^{3(\widetilde{N}_{c} + 1) - R_{f}}}{\Pf (N)}\right )^{1/\widetilde{N}_{c} + 1 - R_{f}}
    +\frac{1}{2}\Tr \widetilde{M}N
\end{align}
where N is the meson field formed by the remaining massless dual quarks. Note that we ignored the $\omega_{\widetilde{N}_{c}+1 - R_f}$ factor in the first term and the Yukawa coupling $\lambda$ in the second term. The second term comes from the Yukawa interaction of the dual theory.

First we consider only the effect of the dominant SUSY potential. Finding the minimum of the SUSY potential is equivalent to solving the SUSY equations of motion (EOM), which set $W$ to the minimum. The EOM of $\widetilde{M}$ sets $N=0$. Meanwhile, the EOM of $N$ is singular at this point, indicating $\widetilde{M}\rightarrow \infty$. This violates the assumption of small $\widetilde M$. Therefore, even before a small AMSB deformation can be applied,
this branch collapses back to the mesonic branch already
considered. \\

For $R_f= \widetilde{N}_{c}+1$, there will be emergent meson degrees of freedom $N$ with a quantum modified constraint. The superpotential would be 
\begin{equation}
    W = \frac{1}{2}\Tr \widetilde M N + X ( \Pf(N) - 2^{\widetilde{N}_c -1} \Lambda_{L}^{2(\widetilde{N}_c +1)})
\end{equation}
where $X$ is a Lagrange multiplier whose SUSY EOM implements the quantum modified constraint. The SUSY potential is non-vanishing and affected by higher order terms in the K\"ahler potential. Notice that the SUSY EOM of $\widetilde M$ sets $N = 0$, whereas the quantum constraint imposes $N = \mathcal{O}(\Lambda_{L}) \neq 0$. 
This behavior suggests that the energy density scale goes back to $\mathcal{O}(\Lambda_L)$.
Like the previous case, this branch also collapses back to the mesonic branch before any AMSB deformation can be applied. \\

For $\widetilde{N}_{c}+2 \leq R_f < 3(\widetilde{N}_{c}+1)$, the IR dynamics of the remaining theory is described by a magnetic dual with gauge group $Sp (R_f - \widetilde N_{c} -2)$ (except for $R_f = \widetilde N_{c} + 2$ where the theory is s-confining). The superpotential is
\begin{equation}
    W_{L} =  \frac{1}{2}N_{ij}b_{ic}b_{jd}J^{cd} + \frac{1}{2}\Tr \widetilde M N
\end{equation}
where the $b$ and $N$ are dual-dual quarks and mesons formed by the remaining massless dual quarks, respectively. The original superpotential term \eqref{eq:W_yukawa} has transformed into the second term of $W_{L}$ above, and this term enforces $N = 0$ as the equation of motion for $\widetilde M$ in the supersymmetric limit. This means when we introduce tree-level AMSB, $N = \mathcal{O}(m)$, and we were justified in ignoring the s-confining $\Pf(N)$ term as a high power of $m$ (assuming $N$ is even full rank). We rescale the fields by appropriate factors of $\Lambda_{L}$ to make them canonical. Ignoring order one factors we have,
\begin{equation}
    W_{L} = \frac{1}{2}N_{ij}b_{ic}b_{jd}J^{cd} + \frac{1}{2}\Lambda_{L}\Tr \widetilde M N
\end{equation}
Finally we substitute the value of $\Lambda_{L}$ to arrive at 
\begin{equation}
     W_{L} =\frac{1}{2}N_{ij}b_{ic}b_{jd}J^{cd} + \frac{1}{2}(\Pf(\widehat M))^{1/(3(\widetilde N_{c} + 1) - R_f)}\Tr \widetilde M N \label{eq:dualdualW}
\end{equation}
Let all fields be real (one can check that minimum can indeed be found under this condition) and consider the direction given by $N = N^{0} \otimes i\sigma_{2}$ with $N^{0}_{ii} = n_{i}$, $\widetilde M = \widetilde M^{0} \otimes i\sigma_2$ with $\widetilde M^{0}_{ii} = x_{i}$, $b = b^{0} \otimes \textbf{1}_{2}$ with $b^{0}_{ii} = y_{i}$, for $i = 1$, $\cdots$, $(R_f - \widetilde N_{c}-2)$ and with all other entries $0$. Finally let $\widehat M = \widehat M^{0} \otimes i\sigma_{2}$ with $\widehat M^{0} = v \textbf{1}$. In this direction all fields, baryonic and mesonic, are turned on. Then by our earlier assumptions, $x_{i} \ll v$, and the SUSY $+$ AMSB scalar potential is 
\begin{align}
   V=&\sum_{i} 4y_{i}^{2}n_{i}^{2} + (v^{C}x_{i} - y_{i}^{2})^{2} + v^{2C}n_{i}^{2} -
   2(C-1)mv^{C}n_{i} x_{i} \nonumber \\
   &\hspace{15pt}+ \frac{C}{3(\widetilde N_{c} + 1) - R_f}v^{2C-2}\left ( \sum_{i} n_{i}x_{i}\right )^{2}
\end{align}
where $C = (N_f - R_f)/(3(\widetilde N_{c} + 1) - R_f)$ and remains greater than 1. Notice that the final term is smaller than the third term in the first sum by a factor of $x_{i}^{2}/v^{2} \ll 1$. Therefore, we can neglect this term and the potential splits into $R_f - \widetilde N_{c}-2$ identical parts. In what follows, we suppress the index $i$. Substituting the $y$ and $n$ equations of motion, and using
$n$, $x \ll \Lambda_{L} = v^{C}$ (the $v$ here has a modified dimension as a consequence of setting $\widetilde{\Lambda} = 1$) along the way, we get
\begin{equation}
    V |_{y,  n} = -(C-1)^{2}m^{2}x^{2}
\end{equation}
This function of $x$ does not have a minimum, but this tree-level runaway is power suppressed as $\mathcal{O}(x^{2}) \ll \mathcal{O}(v^{2})$. Due to the Yukawa term in \eqref{eq:dualdualW}, the dual-dual quarks will acquire a mass $v$, so we can integrate them out. Then AMSB effects yield a 2-loop mass for the meson $\widehat M$, and the corresponding 2-loop potential gives a positive contribution with $\mathcal{O}(v^{2})$, which will stabilize the runaway. However, notice that the resultant minimum keeps getting deeper as $x$ increases. So for enhanced stability, the theory would push $x$ away from the origin until it hits the scale $\Lambda_{L}$. So our initial assumption of the $\widetilde M$ VEVs being much smaller than the $\widehat M$ VEVs falls apart, and we once again collapse to the mesonic branch (where all the meson VEVs are equal). 

\section{Conformal Window $\frac{3}{2} (N_{c}+1) < N_{f} < 3(N_{c} + 1)$}
Seiberg established the conformal window of SQCD for $\frac{3}{2} (N_{c}+1) < N_f < 3 (N_{c}+1)$, where the theory flows to IR fixed points with non-trivial superconformal dynamics. The $Sp(N_{c})$ electric theory (2$N_f$ quarks $Q$ in the fundamental representation) and the $Sp(N_f-N_{c} - 2) = Sp(\widetilde{N}_c)$ magnetic theory (2$N_f$ dual quarks $q$ together with a gauge singlet meson field $M$) are supposed to describe the same physics in the infrared (IR). We assume that the equivalence persists sufficiently near the IR fixed point, and take $m \ll \Lambda, \Lambda_m$ to justify this assumption, where $\Lambda$ ($\Lambda_m$) is the strong scale of the electric (magnetic) theory. The electric theory has no superpotential, while the magnetic theory has
\begin{align}
	W = \frac{1}{4\mu_{\rm m}} M_{ij} q^{i}_{c} q^{j}_{d}J^{cd}
\end{align}
with $M^{ij}$ being dual to the mesons in the electric theory. Here, $\mu_{\rm m}$ is the matching scale that satisfies 
\begin{equation}
    16\mu_{\rm m}^{N_f}(-1)^{N_f - N_{c} - 1} = \Lambda^{3(N_{c} +1)-N_f} \Lambda_m^{3(\widetilde{N}_c+1)-N_f}
\end{equation} In either description, the theory has a global $SU(2N_f)_{Q,q} \times U(1)_R$ symmetry.\\

For a superconformal theory, the conformal dimensions of chiral fields are determined completely \footnote{In general there can be other, non-anomalous $U(1)$ symmetries that make the choice of $U(1)_R$ charges
ambiguous. The correct prescription is given by a-maximization \cite{Intriligator:2003jj}. In this work, only the quarks are charged under the gauge group so the NSVZ $\beta$-function is sufficient to determine the fixed-point anomalous dimension of the quarks. That of the singlet (when present) can then be determined from the running of the Yukawa coupling. For this reason we do not need to invoke a-maximization.} by their $R$-charges, $D(\phi)=\frac{3}{2}R(\phi)$. On the other hand, the $U(1)_R$ symmetry in supersymmetric QCD is determined by the anomaly-free condition and the charge conjugation invariance,
\begin{align}
	R(Q)&= \frac{N_f-N_{c}-1}{N_f}, \nonumber \\
	R(q) &= \frac{N_f-\widetilde{N}_c-1}{N_f}, \qquad
	R(M) = 2\frac{\widetilde{N}_c +1}{N_f}\ . \label{eq:Rcharges}
\end{align}
The anomalous dimension of mass is $2(1-D(\phi))$, and therefore the K\"ahler potential receives the wave function renormalization
\begin{align}
	K = Z_\phi(\mu)\phi^* \phi = \left( \frac{\mu}{\Lambda} \right)^{2-2D(\phi)} \phi^* \phi.
	\label{eq:anomalous}
\end{align}

Here, $\mu$ is the renormalization scale, and $\Lambda$ is the energy scale where theory becomes nearly superconformal.

It is clear that AMSB effects asymptotically vanish towards the IR limit because couplings no longer run,
\begin{align}
	m_{Q,q,M}^2(\mu) \rightarrow 0, &\qquad
	m_\lambda(\mu) \rightarrow 0.
\end{align}
However, the effects are relevant and change the IR dynamics unless
\begin{align}
	\frac{m_{Q,q,M}^2(\mu)}{\mu^2} \rightarrow 0, &\qquad
	\frac{m_\lambda(\mu)}{\mu} \rightarrow 0 \label{eq:amsbrel}.
\end{align}

That is, if the AMSB effects scale sufficiently slowly as $\mu\to0$, they can produce different IR dynamics from the ordinary SQCD case. This procedure and definition of relevance are standard, and can be seen in e.g. section 1.1 of \cite{Rychkov_2017}.

Unfortunately we do not have computational tools to answer this question for the entire range of the conformal window. Instead, we look at Banks--Zaks (BZ) fixed points \cite{Banks:1981nn, Caswell:1974gg} where the conformal dynamics can be studied using perturbation theory. This is possible at the upper edge or the lower edge of the conformal window as the IR fixed point couplings turn out to be perturbative. In the conformal window, the magnetic description is no longer IR free, and has a non-trivial fixed point which is weakly coupled at the lower end of the window. We will first analyze the behavior of AMSB in this region and
then we will turn to the upper end of the window where the electric theory has a weakly coupled fixed point. There will be no runaways in either case; AMSB effects make a relevant deformation in both cases and destroys the superconformal phase. We can only conjecture about the intermediate region where both descriptions are strongly coupled. Finally, we demonstrate local chiral symmetry breaking minima in both regimes and conjecture via interpolation that chiral symmetry is broken throughout the conformal window. 

\subsection{Lower Conformal Window}
We begin by considering the magnetic BZ fixed points for $N_f = 3(\widetilde{N}_c +1)/(1+\epsilon)$, $0 < \epsilon \ll 1$. This is at the lower edge of the conformal window because $N_f \approx 3(\widetilde{N}_c + 1) = 3(N_f - N_{c} - 1)$ implies $N_f \approx \frac{3}{2}(N_{c} + 1)$. We will work in the large $\widetilde N_{c}$ limit and leading non-trivial order in $\epsilon$ for simplicity. For notational
convenience, we define
\begin{align}
	x &\equiv \frac{\widetilde{N}_c}{8\pi^2} \lambda^2, \qquad
	y \equiv \frac{\widetilde{N}_c}{8\pi^2} g^2.
\end{align}

Naively, it would appear that the dynamics is ambiguous because depending on the initial condition of coupling constants, $m_q^2$ and $m_M^2$ are found to have either sign. We will also find that in the deep IR, the coupled beta functions of the gauge coupling $g$ and Yukawa coupling $\lambda$ make them run asymptotically to the IR attractor along a trajectory given by
\begin{equation}
    \frac{d}{d \log \mu}\frac{g^{2}}{\lambda^{2}} = 0 \implies \frac{\beta_{g}}{g^{2}} = \frac{\beta_{\lambda}}{\lambda^{2}} \label{eq:trajectory}
\end{equation}
In other words, as the theory flows to the IR, $x$ and $y$ will approach the fixed point along this specific trajectory, from above or below (depending on initial conditions).   

\begin{figure}[t]
\includegraphics[width=0.8\columnwidth]{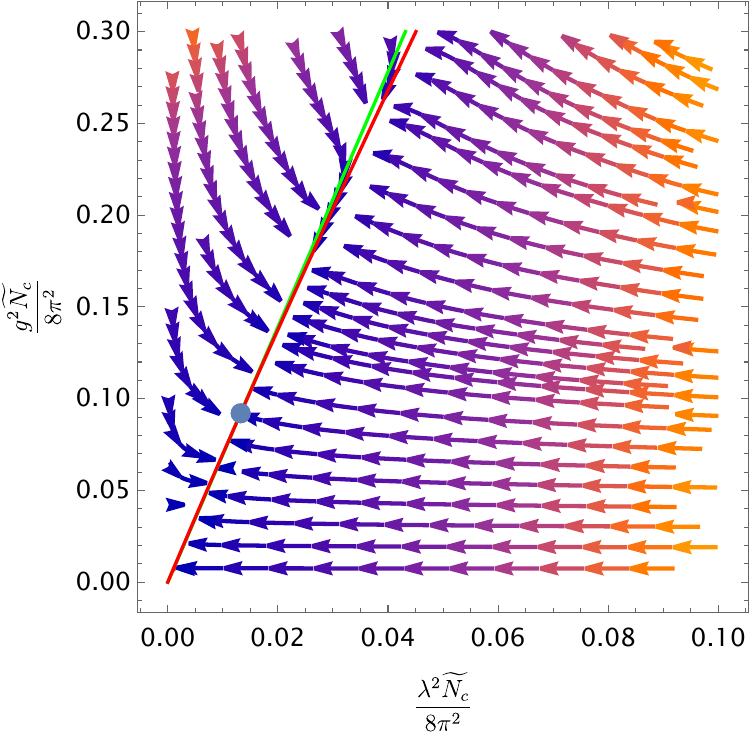}
\caption{Two-dimensional RGE flow of coupling constants near the magnetic Banks--Zaks fixed point (blue dot) with $N_f = 151$, $\widetilde{N}_c=50$. $m_{\widetilde{q}}^2 > 0$ below the red line, while $m_M^2 > 0$ below the green line. }
\label{fig:2DRG}
\end{figure}

The magnetic RGEs modified for our $Sp(\widetilde N_{c})$ gauge theory are given by equations \eqref{eq:anomalous_q} to \eqref{eq:beta_lambda^2_exact}. 
A numerical plot of this RG flow is shown in Figure \ref{fig:2DRG}, superposed with curves of $m_q^2 = 0$ (red) and $m_M^2 = 0$ (green). Notice that irrespective of the trajectories far from the fixed point, the coupling constants always approach the fixed point along a trajectory that lies between the red and green lines.  \\

Taking $\widetilde{N}_c \gg 1$ and expanding in $\epsilon$ to linear order gives linearized RGEs:
\begin{align}
	\beta (y) &= \mu \frac{d}{d\mu} y = -3y^2 ( \epsilon -  y + 6x), \label{eq:RGElambda}\\
	\beta(x) &= \mu \frac{d}{d\mu} x = x ( -2y + 14x ). \label{eq:RGEg} 
\end{align}

By setting the LHS of equations \eqref{eq:RGElambda} and \eqref{eq:RGEg} to zero, one can find that the BZ fixed point is at $(x_{0},y_{0})\approx(\epsilon,7\epsilon)$. The approximation is valid as long as $7\epsilon \ll 1$. 
Define $\delta x = x- x_{0}$ and $\delta y = y- y_{0}$. We study the case where we are close to the fixed point, namely, $\delta x$, $\delta y \ll \epsilon$. \\

If we expand the RGEs around the fixed point we obtain a coupled set of linear first-order differential equations. To leading order in $\epsilon$ the eigensolutions are
\begin{align}
    \frac{d}{d\log\mu}(63\epsilon\delta x + \delta y) \approx& 21\epsilon^2 (63\epsilon\delta x + \delta y)
    \\
    \frac{d}{d\log\mu}(-7(1-\frac{3}{2}\epsilon)\delta x + \delta y) \approx& 14\epsilon (-7(1-\frac{3}{2}\epsilon)\delta x + \delta y)
\end{align}

Then since $21\epsilon^2 \ll 14\epsilon$ the latter combination quickly runs to zero, leaving the remainder of the flow with the slower eigenvalue to occur along the line:
\begin{align}
    \delta x =\frac{1}{7}& \delta y \frac{1}{1-\frac{3}{2}\epsilon} \approx \frac{1}{7}\left(1+\frac{3\epsilon}{2}\right)\delta y
\end{align}

This is consistent with the constraint \eqref{eq:trajectory}. Using this result to express $\beta (x)$, $\beta(y)$ in terms of $\delta y$ and $\epsilon$, we get the RG flow
\begin{align}
    \beta (y) &= 21\epsilon^{2}\delta y \\
    \beta(x) &= 3\epsilon^{2}\delta y \approx 21\epsilon^{2}\delta x 
\end{align}
yielding 
\begin{align}
    \delta y \sim \mu^{21\epsilon^{2}} \\
    \delta x \sim \mu^{21\epsilon^{2}} \label{eq:delyscale}
\end{align}

Therefore, both the couplings approach the fixed point with the exponent $e^{21\epsilon^2 t}$, where $t=\ln \mu$. Notice that these exponentials are slower than $\mu^2 = e^{2t}$ whenever $21\epsilon^2 < 2$, which is indeed the case if $7\epsilon \ll 1$. This satisfies \eqref{eq:amsbrel}. Therefore the AMSB effects are relevant for sufficiently small $\epsilon$. \\

\begin{figure}[t]
\includegraphics[width=0.48\columnwidth]{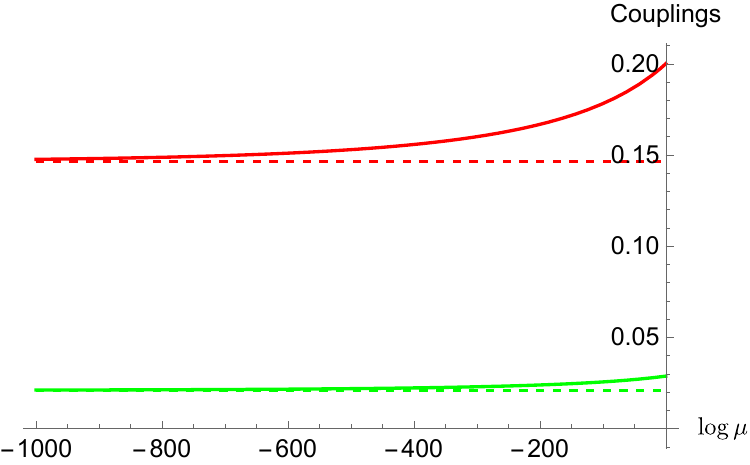}
\includegraphics[width=0.48\columnwidth]{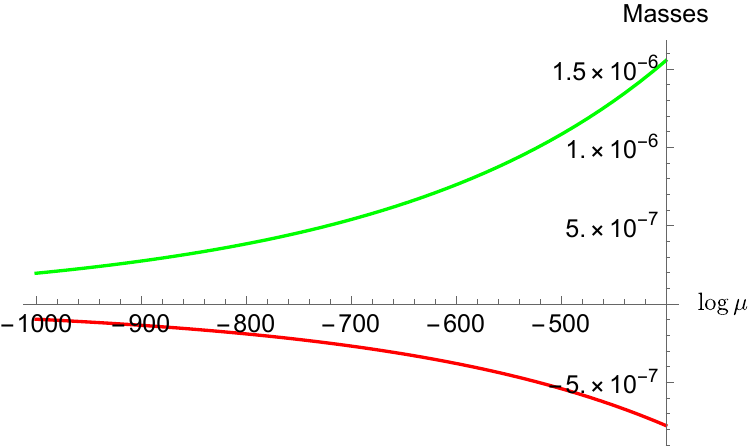}\\
\caption{Above: Running of $g^2$ (red) and $\lambda^2$ (green),  
with the initial condition $g^2 = 0.2$, $\lambda^2 = \frac{1}{7} g^2 + 0.00015$, while the dashed lines show their infrared fixed point values. We took $N_{c} = 99$, $N_f=151$, $\widetilde{N}_c=50$, and hence $\epsilon \approx 0.0132$. 
Below: Corresponding running of $m_q^2$ (red) and $m_M^2$ (green) in units with $m=1$.}
\label{fig:runmass}
\end{figure}

Using \eqref{eq:m2_anomaly_scalar}, the meson and dual squark masses are

\begin{align}
    m_{M}^{2} &= \frac{3}{2}\epsilon^{2} \delta y m^{2} \\
    m_{q}^{2} &= -\frac{3}{4}\epsilon^{2}\delta y m^{2} \label{eq:massesE}
\end{align}
Notice that if we approach the fixed point from below in coupling space, then $\delta y < 0$ and hence $m^{2}_{M} < 0$, $m^{2}_{q} > 0$. On the other hand, if we approach the fixed point from above in coupling space, then $\delta y > 0$ and hence $m^{2}_{M} > 0$, $m^{2}_{q} < 0$. An illustration of the running of coupling constants, and its consequences for the squark and meson masses near the infrared magentic Banks-Zaks fixed point, is shown in Figure \ref{fig:runmass}, where we have picked initial conditions such that the couplings approach the fixed point from below. For numerical plots, we took values already close to the IR fixed point to keep the amount of running manageable. 

Pushing $\epsilon$ beyond the validity range, we see that AMSB effects are relevant if $21\epsilon^2 < 2$, or equivalently $N_f < 1.77 N_{c}$. But we cannot trust this upper bound given that the approximations made by us are no longer valid there. While we cannot exclude the possibility that AMSB is irrelevant for $N_f \sim 1.7 N_{c}$, we will now derive chiral symmetry breaking vacua and show a consistent picture that it is likely relevant. 

Notice that if we pick the initial conditions such that $\delta y < 0$, the dual squarks are tachyonic. However, this does not lead to a runaway due to the presence of a positive quartic term in the potential, as we shall see below.

\subsubsection{Branch with Dual Squark VEVs in the Magnetic Theory}
 Consider the low energy theory obtained from initial conditions with $m_q^2<0$, where we give D-flat VEVs to the dual squark. For these initial conditions the squark VEV appears to trigger condensation of the mesons as well, leading to a minimum along a mixed branch. We focus on the slice of moduli space parametrized by the VEVs $q = \phi_q \textbf{1}\otimes \textbf{1}_2$ and $M = \phi_M \textbf{1}\otimes i\sigma_2$. \footnote{If we just used the available gauge/flavor rotations and D-flat conditions, generally there could be more non-zero entries in the $q$ matrix, whose presence in the scalar potential could be parametrized using a CKM-esque unitary matrix. However, even a numerical exploration turns out to be very challenging if we assume full generality. Therefore, we resort to a numerical exploration on a slice of enhanced symmetry.} The superpotential and K\"ahler potential are then
 
 \begin{align}
	W = \frac{\lambda}{2} M_{ij} q^{i}_{c} q^{j}_{d}J^{cd} = \lambda \widetilde{N}_{c}\phi_M \phi_q^{2}\label{eq:W_yukawa}
\end{align}
\begin{align}
	K = c_M\frac{Z_M}{2}\Tr M^{\dagger}M + c_q Z_q\Tr q^{\dagger}q = c_M N_f Z_M \left|\phi_M \right|^{2} + 2c_q \widetilde{N}_{c}Z_q\left|\phi_q \right|^{2}
\end{align}
which lead to the scalar potential 
\begin{align}
  V&=  -c_MN_fZ_M(\mu)m^2\frac{\dot{\gamma}_M}{4} |\phi_M|^2
+\frac{1}{c_MN_fZ_M(\mu)} \left| \frac{\partial W}{\partial\phi_M} \right|^2
- 2c_q\widetilde{N}_{c}Z_q(\mu)m^2\frac{\dot{\gamma}_q}{4} |\phi_q|^2
\nonumber\\
+&\frac{1}{2c_q\widetilde{N}_{c}Z_q(\mu)} \left| \frac{\partial W}{\partial\phi_q} \right|^2 
+3m(W+h.c.) -m \left( 1-\frac{\gamma_M}{2} \right)\left(\phi_M\frac{\partial W }{\partial \phi_M }+h.c.\right) \nonumber\\
-&m \left( 1-\frac{\gamma_q}{2} \right)\left(\phi_q\frac{\partial W }{\partial \phi_q }+h.c.\right)  \nonumber\\
&=  -c_MN_fZ_M(\mu)m^2\frac{\dot{\gamma}_M}{4} \phi_M^2
+\frac{(\lambda\widetilde{N}_{c} \phi_q^{2})^2}{c_MN_fZ_M(\mu)}
- 2c_q\widetilde{N}_{c}Z_q(\mu)m^2\frac{\dot{\gamma}_q}{4} \phi_q^2
\nonumber\\
+&\frac{(2\lambda\widetilde{N}_{c}\phi_M \phi_q)^2}{2c_q\widetilde{N}_{c}Z_q(\mu)} +m \left(\gamma_M + 2\gamma_q\right)\left(\lambda\widetilde{N}_{c}\phi_M \phi_q^{2}\right)
\end{align}

Recall that if we are close to the lower edge of the conformal window, 
\begin{align}
    \beta (y) &= 21\epsilon^{2}\delta y \implies y(t) = 7\epsilon + (y(0) - 7\epsilon)e^{21\epsilon^2 t} \\
    \beta(x) &= 21\epsilon^{2}\delta x \implies x(t) = \epsilon + (x(0) - \epsilon)e^{21\epsilon^2 t}
\end{align}
Hence the anomalous dimensions, if $x, y \sim \epsilon$, are
\begin{align}
  \gamma_q &= 
  \epsilon + (y(0) - 6x(0) - \epsilon)e^{21\epsilon^2 t} \\
  \gamma_M &= 
  -2\epsilon -(2x(0) - 2\epsilon)e^{21\epsilon^2 t}
\end{align}

\begin{figure}[h]
\includegraphics[width=0.92\columnwidth]{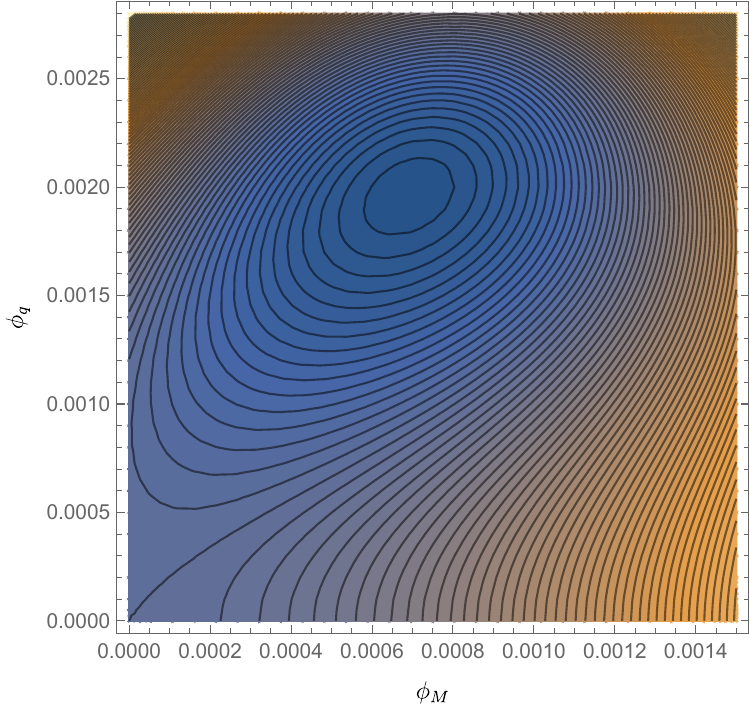}\\
\caption{A contour plot of the scalar potential in the lower conformal window when the dual squarks are tachyonic and the meson mass-squared is positive. Both the dual squark VEV $\phi_q$ and the meson VEV $\phi_M$ are non-zero at the unique minimum. For this plot, we have assigned the numerical values $\widetilde{N_c} = 10$, $\epsilon = 0.01$, $\frac{m}{\Lambda_m} = 0.1$, $c_M = c_q = 1$, $x(0) = 0.03$ and $y(0) = 0.2$. The value of the scalar potential at the minimum is $-6.53884 \times 10^{-12}$ in units where $\widetilde{\Lambda_m} = 1$.}
\label{fig:CWContour}
\end{figure}

Note that $Z_M = \left( \frac{\mu}{\Lambda_m}\right)^{-2\epsilon}$, $Z_q = \left( \frac{\mu}{\Lambda_m}\right)^{\epsilon}$ and $N_f = 3(\widetilde{N}_c + 1)/(1+\epsilon)$. Even for the slice that we have chosen, the scalar potential is complicated and an analytical derivation of the minimum does not seem to be possible. The scalar potential is numerically analyzed using contour plots on the $\phi_M-\phi_q$ plane, for different values of the "unconstrained" parameters $\widetilde{N_c}$, $\epsilon$, $\frac{m}{\Lambda}$, $c_M$, $c_q$, and the couplings $\{x(0), y(0)\}$ in the UV. \footnote{These initial conditions for coupling values are taken assuming flow to the fixed point from ABOVE in coupling space, which is required for tachyonic dual squarks and positive meson mass-squared, and hence the possibility of a finite minimum along the mixed q \& M branch.} We take $\mu \approx \frac{1}{2}(\phi_M + \phi_q)$ (one could also take the geometric mean without changing the net qualitative outcome of the analysis). Regardless of these choices, we always find a single minimum at non-zero $\phi_M$ and $\phi_q$ (hence the "mixed branch"). In particular, there is no runaway despite tachyonic dual squarks, and also no minimum is seen along the pure dual squark branch (i.e. for $\phi_M = 0$ and $\phi_q \neq 0$). For illustration, one contour plot has been shown in Figure \ref{fig:CWContour} along with the corresponding numerical choices made. \\

We will also show the existence of pure mesonic branch in co-existence with this mixed branch at the end of the next subsection, but first we shall look at the complementary case where initial conditions are such that $m_M^2 = -m^2\frac{\dot{\gamma}_M}{4} < 0$ and $m_q^2 = -m^2\frac{\dot{\gamma}_q}{4} > 0$. Notice that in this case the scalar potential term quadratic in $\phi_M$ would be negative and there is no quartic term in $\phi_M$ to stabilize the resulting runaway. Hence there would be no finite minimum along a mixed branch where both $\phi_M$ and $\phi_q$ are non-zero. Also, on taking a full rank meson VEV, all the scalar potential terms that are quartic and quadratic in $\phi_q$ would be positive (the sign of the term linear in $m$ could be flipped by doing a $U(1)_R$ rotation on $m$), allowing us to integrate out the dual squarks by taking $\phi_q = 0$.  

\subsubsection{Purely Mesonic Branch in the Magnetic Theory}
As a consequence of a finite full-rank meson VEV, just as in the free magnetic phase, the massive dual quarks can be integrated out and without their effects, the beta function of the gauge theory flips sign, allowing the theory to generate a new IR dynamical (magnetic) scale $\Lambda_m$. The low-energy pure SYM develops a gaugino condensate with the {\it exact}\/ non-perturbative superpotential
\begin{align}
W &= (\widetilde N_{c} +1)\left(2^{\tilde{N}_c-1}\Lambda_m ^{3(\widetilde{N}_{c}+1)-N_{f}} \Pf \left(\frac{M}{2\mu_m}\right)\right)^{1/(\widetilde N_{c}+1)}
\end{align}
Here $[M] = 2$ and it is related to the dual Meson $M_m$ via the relation $M = 2\lambda \mu_m M_m$, so that we can write $\frac{\lambda}{2}M_{mij}q^{c}_{i}q^{d}_{j}J^{cd} = \frac{1}{4\mu_m}M_{ij}q^{c}_{i}q^{d}_{j}J^{cd}$. Taking VEVs $M_{ij} = 2\lambda \mu_{\rm m} \phi \delta_{ij} \otimes i\sigma_2$, we find
\begin{align}
W &= (\widetilde N_{c} +1)\left(2^{N_{f}-N_{c}-3}\Lambda_m ^{3(\widetilde{N}_{c}+1)-N_{f}}\lambda^{N_f} \phi^{N_f}\right)^{1/(\widetilde N_{c}+1)} \nonumber \\
K &= c_M \frac{Z_M}{2}\Tr M_m^{\dagger}M_m = c_M N_f Z_M \phi^{\dagger}\phi \nonumber \\
V &= -c_MN_fZ_M(\mu)m^2\frac{\dot{\gamma}_M}{4} \phi^2 + \frac{2^{2(\widetilde{N}_c-1)/(\widetilde{N}_c + 1)}}{c_MN_fZ_M(\mu)}N_{f}^2
	 \left(\Lambda_m^{3(\widetilde{N}_{c} +1)-N_{f}}\lambda^{N_{f}} \phi^{N_{f}-\widetilde N_{c} -1} \right)^{\frac{2}{\widetilde{N}_c + 1}} \nonumber \\ &-2^{2\widetilde{N}_c/(\widetilde{N}_c + 1)}m \left( 3 (\widetilde{N}_c + 1) - N_f + N_f\frac{\gamma_M}{2}\right) 
		\Lambda_m^{3-\frac{N_f}{\widetilde N_{c}+1}} \left(\lambda \phi \right)^{N_{f}/(\widetilde{N}_c + 1)}
\end{align}
Taking canonical normalization $\phi^2 \rightarrow \phi^2 / (N_f c_M Z_M(\mu))$, using $\gamma_M = 2\frac{Nf - 3(\widetilde{N}_{c}+1)}{Nf} + a\left(\frac{\mu}{\Lambda_m}\right)^{\alpha}$, $Z_M = \left( \frac{\mu}{\Lambda_m}\right)^{-2\epsilon}$, $N_f = 3(\widetilde{N}_c + 1)/(1+\epsilon)$, $\widetilde{N}_c \gg 1$, and finally setting $\mu \approx \phi$, we arrive at
\begin{align}
   \frac{V}{\Lambda_m^{4}} = &-\frac{21\epsilon^2 a}{4}\left(\frac{m}{\Lambda_m}\right)^2\left( \frac{\phi}{\Lambda_m}\right)^{2+21\epsilon^2} + 4N_f^2\left(\frac{\lambda}{\sqrt{c_M Nf}}\right)^{6(1-\epsilon+\epsilon^2)}\left( \frac{\phi}{\Lambda_m}\right)^{4} \nonumber \\
    &-4a N_f \left(\frac{m}{\Lambda_m}\right) \left(\frac{\lambda}{\sqrt{c_M Nf}}\right)^{3(1-\epsilon+\epsilon^2)} \left( \frac{\phi}{\Lambda_m}\right)^{3+21\epsilon^2}
\end{align}
If we minimize the most relevant part of the potential, which for $\phi \sim m$, $\alpha = 21\epsilon^2$ and $\epsilon\ll 1$ includes only the first two terms in the RHS above, we find
\begin{align}\label{eq:solmag}
    \frac{\phi}{\Lambda_m} =& \left(\frac{a c_M^{\frac{3}{1+\epsilon}} \alpha(2+\alpha) m^2}{64 \lambda^{\frac{6}{1+\epsilon}} N_f^{2 - \frac{3}{1+\epsilon}} \Lambda_m^2}\right)^{\frac{1}{2-\alpha}} \propto \left(\frac{m}{\Lambda_m}\right)^{\frac{2}{2-\alpha}} \approx \left(\frac{m}{\Lambda_m}\right)^{1+\frac{21\epsilon^2}{2}}
\end{align}
Even if we had an analytical expression for the mixed branch minimum from the previous subsection, the two minima are not to be compared as they are obtained by assuming different sets of initial conditions on the couplings (and hence there is no notion of tunneling between the two). \\

Let us now return to the case where the initial conditions are such that $m_q^2 < 0$ and $m_M^2 > 0$. Recall that we had identical conditions in the upper end of the free magnetic phase, where it was shown that a pure mesonic branch minimum co-exists with a mixed $q$ \& $M$ branch minimum. Following a very similar procedure here, we will provide {\it a posteriori}\/ justification for integrating out the dual squarks. Continuing the example used in Figure \ref{fig:CWContour}, let us discuss the case where $\widetilde{N_c} = 10$, $\epsilon = 0.01$, $\frac{m}{\widetilde{\Lambda}} = 0.1$, $c_M = c_q = 1$, $x(0) = 0.03$ and $y(0) = 0.2$. Note that in the result (6.29), $a = (2\epsilon - 2x(0)$ if we follow equation (6.25) and we substitute for $\lambda$ from equation (6.22), taking the RGE scale $\mu \approx \phi_M$. Then the minimum is found to be at $\phi = 0.133576$, in units of $\Lambda_m = 1$. So we plug in $\phi_M = 0.133576$ into equation (6.22), and follow the steps in the previous subsection to express this effective scalar potential as a function of $\phi_q$ only. We find that $\phi_q = 0$ is indeed a minimum and hence conclude that it was justified to integrate out the dual squarks in the first place. It is also found that pure mesonic minimum is deeper than the mixed $q$ \& $M$ minimum. In this example, the pure mesonic potential has a minimum value of $-2.0966\times 10^{-7}$ (in units of $\Lambda_m = 1$), which is much deeper than the mixed brach potential's minimum value of $-6.53884 \times 10^{-12}$. \\

Based on many numerical explorations similar to the example presented above, going up to $\widetilde{N_c} = 15$, it seems very convincing that the pure mesonic branch provides the global (fully) chiral symmetry breaking minimum at the lower edge of the conformal, irrespective of whether we flow to the fixed point from above or from below in coupling space. At this minimum, the original $SU(2N_f)$ chiral flavor symmetry is completely broken into $Sp(N_f)$ by meson VEVs. In addition to this global minimum, the mixed $q$ and $M$ branch features a local (partial) chiral symmetry breaking minimum when we flow to the fixed point from above, where the original $SU(2N_f)$ chiral flavor symmetry is broken into $Sp(\widetilde{N_c}) \otimes Sp(N_f - \widetilde{N_c} -2)$.

\subsection{Upper Conformal Window}
In this section, we show that the AMSB effects are relevant and modify the IR dynamics at the higher edge of the conformal window, using the electric BZ fixed point for $N_f = 3(N_{c}+1)/(1+\epsilon)$, $0 < \epsilon \ll 1$. Again we work in the large $N_{c}$ limit and leading non-trivial order in $\epsilon$. We will begin with the electric description, but then for deriving the chiral symmetry breaking minimum we will have to resort to the twice-dual theory. 

\subsubsection{RG in The Electric Theory}
\begin{figure}[t]
\centering
\includegraphics[width=0.61\columnwidth]{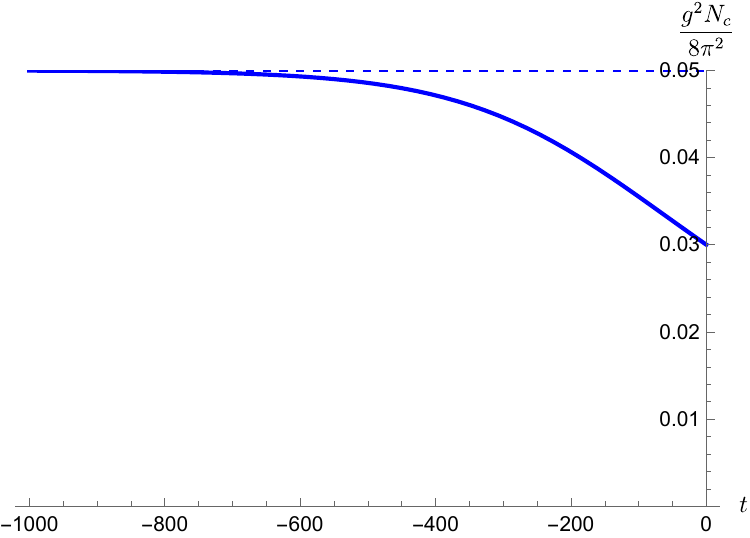}\\
\includegraphics[width=0.61\columnwidth]{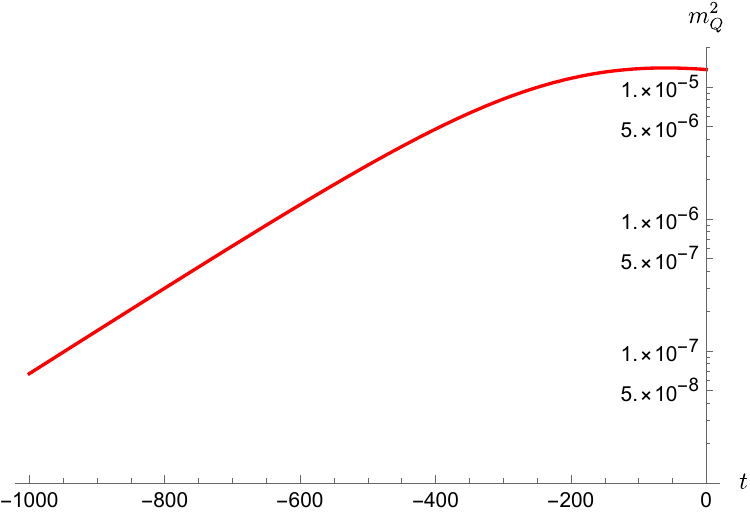}\\
\caption{Running of the electric gauge coupling (above) and the AMSB quark mass squared (below) in units with $m=1$ near the electric Banks--Zaks fixed point with $\epsilon=1/20$ and $N_{c} g^2(0)/8\pi^2 = 0.03$ and $N_c \gg 1$. The dashed line represents the infrared fixed point.}
\label{fig:running}
\end{figure}

Running effects in the electric theory are given by \cite{MartinVaughn}
\begin{align}
	\gamma_Q &= \frac{1}{8\pi^2} (2g^2 C_F), \\
	\beta(g) &= \mu\frac{d}{d\mu}g^2 
	= -g^4 \frac{3(N_{c} +1) - N_f - N_f \gamma_Q}{8\pi^2 - (N_{c} +1) g^2}\ ,
\end{align}
with $C_F = (2N_{c} + 1)/4$. 
For $N_f = 3(N_{c} +1)/(1+\epsilon)$, $N_{c} \gg 1$ (i.e. $N_{c} +1 \approx N_{c})$, and $y \equiv N_{c} g^2/ 8\pi^2 $, the beta function, squark and gluino masses reduce to 
\begin{align}
	\mu \frac{d}{d\mu} y &= -3y^2 ( \epsilon - y), \\
	m_Q^2 &= \frac{3}{4} y^2 (\epsilon-y) m^2,  \\
    m_{\lambda} &= \frac{3}{2}(\epsilon-y)m
\end{align}
at the leading order in $\epsilon$. Note that since the electric theory is weakly coupled in the UV, the couplings approach the fixed point from below as they flow from UV to IR regimes. Therefore, $\epsilon > y$ and the squark mass is positive. The approximations we made are valid when $y \approx \epsilon \ll 1$, i.e. we are in the neighborhood of the fixed point (setting the variation of $y$ w.r.t $\mu$ equal to zero, we see that $y = \epsilon$ at the fixed point). An example of the solutions is shown in Figure~\ref{fig:running}. Note that in the large $N_c$ limit that we are taking, the results agree between this theory and the electric $SU(N_c)$ AMSB theory \cite{Kondo:2021osz}.\\

We can work out the approximate solution near the fixed point as
\begin{align}
	y(t) - \epsilon &=  (y(0)-\epsilon) e^{3\epsilon^2 t}, 
	\label{eq:y} \\
	m_Q^2 &= \frac{3}{4} \epsilon^2 (\epsilon -y(0)) e^{3\epsilon^2 t} m^2,
	\label{eq:mQ2} \\
 m_{\lambda} &= \frac{3}{2} (\epsilon -y(0)) e^{3\epsilon^2 t} m 
\end{align}
Here $t = \ln \mu \rightarrow -\infty$ defines the IR limit. The AMSB effects are relevant when $3\epsilon^2 < 2$, or equivalently $N_f > 1.65 N_{c}$ (this is when the above exponentials are slower than $\mu^{2} = e^{2t}$). However, we cannot fully trust this lower bound since our approximations of small $\epsilon$ are no longer valid near it. Since $m_Q^2$ stays positive through all energy scales, there is no minimum along this direction. At some point in the RG flow the squark and gluino masses will exceed the renormalization scale. At this point the superpartners decouple,
can be integrated out, and the superconformal phase is
destroyed. What remains is non-SUSY QCD at low energies, which must
be analyzed from the (albeit strongly-coupled) magnetic
description or the twice-dual of the electric theory (see the next subsection). \\

In the vicinity of the upper edge of the conformal window, if we consider giving the squarks in the electric theory a generic D-flat VEV of the form
\begin{equation}
    Q = 
\begin{pmatrix}
 a_{1} &  &  &  &  &  &\\
  & a_{2} &  &  &  &  &\\
  &  & \ddots &  &  &  &\\
  &  &  & a_{N_{c}} &  &  &
\end{pmatrix} \otimes \textbf{1}_{2}
\end{equation}
then the only non-zero contributions to the scalar potential would be in the form of 2-loop AMSB mass terms:
\begin{equation}
    V_{Q} = m_{Q}^{2}\sum^{N_c}_{i=1} |a_{i}|^{2} = \frac{3}{4} y^2 (\epsilon-y) m^2\sum^{N_c}_{i=1} |a_{i}|^{2}
\end{equation}
The minimum would appear to be the origin in the electric description. The low energy theory is then described by integrating out the massive gauginos and scalars. The beta function of the theory turns back over, leading to strong coupling and cutting off our ability to analyze it directly. In the next section, we show how an alternative description can yield results.\\

\subsubsection{The Twice Dual Theory}
Applying the duality map again leads to a $Sp(N_c)$ theory with $2N_f$ quarks in the fundamental representation, along with singlet mesons $M_{ij}$ and $N_{ij}$. The superpotential is
\begin{align}
    W =& \frac{1}{2}Y N_{ij}(Q_{ia}Q_{jb}J^{ab}-\Lambda M_{ij})
\end{align}
In the SUSY theory, the field $N$ serves as a Lagrange multiplier, whose equations of motion enforce the identification of $\Lambda M_{ij}$ with the composite $Q_{ia}Q_{jb}J^{ab}$, leaving us with the original electric theory with $W=0$. However, the presence of AMSB can lead to a nonzero VEV of the field $N$, modifying the story. If, as we have been assuming, Seiberg duality is robust to the AMSB perturbation, then in the deep IR the AMSB-perturbed Twice-Dual description should describe the same physics as the AMSB-perturbed Electric theory.

\subsubsection{RG in The Twice Dual}\label{sec:2DRGE}
Here we analyze the RG equations for the Twice dual theory. Note that the interaction of $N$ with $Q$ is totally analogous to that of the Magnetic theory, while the field $M$ has no interactions that can generate wavefunction renormalization. So the RGE here will be qualitatively the same as that of the Magnetic description.

At 1-loop, the anomalous dimensions are given by
\begin{align}
    \gamma_Q =& \frac{1}{8\pi^2}(2C_Fg^2-(2N_f-1)\lambda^2)
    \\
    \gamma_N =& -\frac{2N_c}{8\pi^2}Y^2
\end{align}
where $C_F = (2N_c+1)/4$ is the quadratic Casimir of $Sp(N_c)$. The beta functions of the gauge and Yukawa couplings are\footnote{There is also RG for the mass term $NM$, but this is just the wavefunction renormalization for $N$. As expected, it flows to zero in the IR.}
\begin{align}
    \dod{g^2}{\ln\mu} =& -\frac{g^4}{8\pi^2}\frac{3(N_c+1)-N_f(1+\gamma_Q)}{1-\frac{(N_c+1)g^2}{8\pi^2}}
    \\
    \dod{Y^2}{\ln\mu} =& -(\gamma_N+2\gamma_Q)Y^2 
\end{align}
We will work in terms of `t Hooft-like couplings $x\equiv\frac{N_cY^2}{8\pi^2}$ and $y\equiv\frac{N_c g^2}{8\pi^2}$ and take $N_f=\frac{3(N_c+1)}{1+\epsilon}$ such that $\epsilon\ll 1$ corresponds to being near the upper edge of the conformal window where the electric description should have a perturbative fixed point. At large $N_c$ and to leading order in $\epsilon$, we have a nontrivial fixed point at
\begin{align}
    (x_*,y_*) =& (\epsilon,7\epsilon)
\end{align}
If we consider small deviations $\delta x$ and $\delta y$ away from this fixed point, and linearize the RGE, we find
\begin{align}
    \dod{}{\ln\mu}\begin{pmatrix}
        \delta x \\ \delta y
    \end{pmatrix} =& \begin{pmatrix}
        14\epsilon & -2\epsilon \\
        -882\epsilon^2 & 147\epsilon^2
    \end{pmatrix}\begin{pmatrix}
        \delta x \\ \delta y
    \end{pmatrix}
\end{align}
The eigenvalues of this matrix are $14\epsilon$ and $21\epsilon^2$. Conclusions about the RG in the Twice-Dual are then completely analogous to those in the Magnetic theory.  As one flows toward the fixed point the larger exponent $14\epsilon$ will decay rapidly, leaving the final approach along the direction corresponding to $21\epsilon^2$. The dynamics is again ambiguous because depending on the initial condition of coupling constants, $m_Q^2$ and $m_N^2$ are found to have either sign. However, our RGE analysis in the electric theory (in section 6.2.1) unambiguously showed that $m_Q^2 > 0$ in the viscinity of the electric Banks-Zaks fixed point, and we expect that in the deep IR the AMSB-perturbed Twice-Dual and Electric theories describe the same physics. So we assume that the initial conditions in the Twice-Dual theory lie in the region of the parameter space where $m_Q^2 > 0$ and $m_N^2 < 0$ (the corresponding flows approach the fixed point from below in coupling space). This would imply that we can integrate out the quarks. \\

\subsubsection{Chiral Symmetry Breaking Minimum in the Twice Dual}\label{sec:2Dminimum}
When $N$ gets a nonzero VEV and the quarks do not, we can integrate the latter out to get an effective superpotential 
\begin{align}
    W =& - \frac{1}{2}Y\Lambda N^{ij}M_{ij} + (N_c+1)\left(2^{N_c-1}\Lambda^{3(N_c+1)-N_f} \Pf{(YN)}\right)^{\frac{1}{N_c+1}}
\end{align}
where $\Lambda M_{ij}$ would be identified with $Q_{ia}Q_{jb}J^{ab}$ in the SUSY limit. We can take $M_{ij} = \delta_{ij}\frac{ \phi_M}{\sqrt{c_M N_f}}\otimes i\sigma_2$ and $N_{ij}=\delta_{ij}\frac{\phi_N}{\sqrt{c_NZ_N N_f}}\otimes i\sigma_2$, such that $\phi_N,\phi_M$ will have canonical K\"ahler potentials. Then
\begin{align}
    W =& Y\Lambda \frac{\phi_N\phi_M}{\sqrt{c_Nc_MZ_N}} + (N_c+1)\left(2^{N_c-1}\Lambda^{3(N_c+1)-N_f} Y^{N_f}\phi_M^{N_f}\right)^{\frac{1}{N_c+1}}
\end{align}
The scalar potential for the canonically normalized $\phi_N,\phi_M$ is then (assuming real moduli for simplicity) 
\begin{align}
    V =& 
    \abs{\dpd{W}{\phi_N}}^2 + \abs{\dpd{W}{\phi_M}}^2
    -\frac{1}{4}\dot{\gamma}_N \phi_N^2 + 2m(1-\frac{\gamma_N}{2}) \phi_N \dpd{W}{\phi_N}
    + 2m \phi_M\dpd{W}{\phi_M} -6mW 
\end{align}
Take $Z_N=(\frac{\mu}{\Lambda})^{\gamma_N}$ and parametrize $\dot{\gamma}_N = c_\gamma(\frac{\mu}{\Lambda})^{\alpha}$. The most relevant part of this potential for small $\phi_N,\phi_M,\mu \ll \Lambda$ is given by (in units with $\Lambda=1$)
\begin{align}
    V =& -\frac{1}{4}c_\gamma \alpha m^2 \phi_N^2 \mu^\alpha + \frac{Y^2}{c_Mc_N}\mu^{-\gamma_N}\phi_N^2 - \frac{4Y}{\sqrt{c_Mc_N}} m\phi_N\phi_M \mu^{-\gamma_N/2} \nonumber\\
    &+\frac{2Y}{\sqrt{c_M c_N}}(1-\frac{\gamma_N}{2}) m \phi_M \phi_N \mu^{-\gamma_N/2} + \frac{Y^2}{c_Mc_N}\phi_M^2 \mu^{-\gamma_N} + \mathcal{O}(\phi_N^x \phi_M^y \mu^z,\, x+y+z \gtrsim 3)
\end{align}
We found from our RGE analysis, see \Cref{sec:2DRGE}, that close to the upper edge of the window we have $\alpha=\mathcal{O}(\epsilon^2)$, where $N_f = \frac{3(N_c+1)}{1+\epsilon}$ and $\epsilon\ll 1$. The first term above, then, is suppressed by a coefficient of order $\epsilon^2$ compared to the others. Sufficiently close to the upper edge, then, it should be consistent to drop it. In that limit, the potential is symmetric (to leading order) in the fields $\phi_N$ and $\phi_M$. This motivates taking $\phi_N=\phi_M\equiv\phi$ and taking $\mu=\phi$ for a Coleman-Weinberg-like potential. Doing so, we obtain a fully chiral symmetry breaking minimum at
\begin{align}
    \frac{\phi}{\Lambda} =& \left[
    \frac{\sqrt{c_Mc_N}(4+2\epsilon)(2-2\epsilon)}{4(2+2\epsilon)Y}\frac{m}{\Lambda}
    \right]^{1/\epsilon} \simeq \left[\frac{\sqrt{c_Mc_N} m}{Y\Lambda}\right]^{1/\epsilon}
    \label{eq:TWDmin}
\end{align}

\begin{figure}[h]
    \centering
    \includegraphics[width=0.8\linewidth]{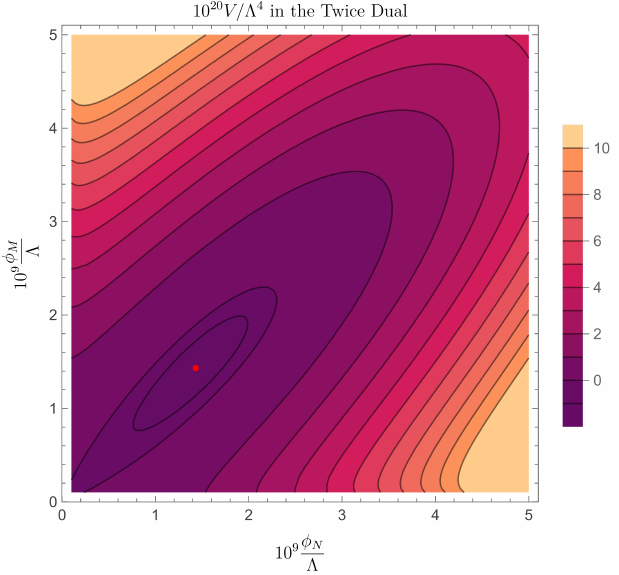}
    \caption{The plot of the potential indicating the minimum at $\phi_M=\phi_N$. We took $N_c=100,N_f=270$, $m=0.1\Lambda,Y=1$, and $\mu=\sqrt{\phi_M\phi_N}$ as an example. The red dot corresponds to the calculated minimum in \eqref{eq:TWDmin}. Note that this is the full potential (with $c_M=c_N=c_\gamma=1$) without any truncation, large $N$ or small $\epsilon$ approximations. Different choices of $N_f,N_c$ give analogous results. Different choices of $\mu$ slightly deform the contours, but keep the minimum very close to $\phi_M=\phi_N$. Everything's been scaled so as to be visible.}
    \label{fig:PotentialTWD}
\end{figure}

We plot the full potential numerically and find the same result, see Figure \ref{fig:PotentialTWD}. This is all identical to the result found for the case of $SU(N_c)$ in \cite{Kondo:2021osz}, which is to be expected since the small $\epsilon$ limit is necessarily large $N_c$. 

\section{Free Electric Phase $N_{f} \geq 3(N_{c}+1)$}
For large number of flavors, the 2-loop squark mass
from AMSB is negative, leading to true runaway behavior. 
AMSB cannot be used to understand the non-SUSY theory in this case.

\section{Conclusions}

We carefully analyzed the behavior of $Sp(N_c)$ gauge theories with $N_f$ flavors upon the application of AMSB, focusing on the chiral symmetry breaking minima and potential baryonic runaway directions. For $N_c + 2 \leq
N_f \leq 3/2(N_c + 1)$ we found that there are no tree level runaways, even though the 2-loop mass of dual quarks turn negative near the upper end of the free magnetic phase, $N_f \gtrsim 1.43N_c$. For similar reasons, runaways are also removed in the lower end of the conformal window.

In summary, we found 
that stable chiral symmetry breaking minima are present for $N_f < 3(N_c + 1)$ upon application of AMSB in the small SUSY-breaking limit. The case of $N_f = N_c + 1$ required particular care due to the inherently strongly coupled nature of the quantum modified moduli space. We found that the theory is best analysed after implementing the quantum constraint, and upon application of AMSB, the chiral symmetry breaking point is found to be stable. Furthermore, the theories with $N_f \gtrsim 1.43N_c$ are protected from runaways to incalculable minima, particularly in the dual quark branch of the free magnetic phase and the lower end of the conformal window. This indicates that the minima that we have found in these regions are indeed the global minima even if we consider large field values of $\mathcal{O}(\Lambda)$. See \autoref{table:phases}
for a case-by-case list of whether we have a global or local minimum for chiral symmetry breaking, and what symmetry breaking pattern there is for the global minimum. The table also includes the results for the corresponding phases of the $SU(N)$ case from \cite{GuideToASQCD} for comparison.  \\

We point out that the case of $N_c=1$ and $N_f=3$ is special. The s-confining superpotential is renormalizable with a dimensionless coupling and has no tree-level AMSB effects. Loop-level effects do not induce chiral symmetry breaking \cite{delima2023sconfining}. How this case is connected to the non-supersymmetric limit $m\rightarrow\infty$ remains an open question.

Our analysis was performed in the $m \ll \Lambda$ limit, and the question remains about the behavior in the non-supersymmetric limit of $m \gg \Lambda$. The chiral symmetry breaking minima for all flavors may or may not be continuously connected to the true vacua of non-SUSY QCD. Irrespective of the potential appearance of a phase transition between these two limits (see arguments based on holomorphy in [22, 23], and also see [34, 35])%
, these are the vacua that are of phenomenological interest for the study of real-world QCD.

\begin{table*}[ht]

\begin{center}
\parbox{13.5cm}{\caption{A comparison between the $\chi_{SB}$ vacua of $Sp(N_c)$ and $SU(N_c)$ gauge theories (note that the latter have a slightly different relation between $N_c$ and $N_f$ for a given phase than what is written in the leftmost column). In each case, if there are both global and local minima in the theory, we first specify the global minimum and the residual flavor symmetry, followed by the local minimum. Note that in the upper edge of the free magnetic phase and at the lower edge of the conformal window, the calculable global minimum in the $Sp(N_{c})$ theory is $\mathcal{O}(-m^{4})$, whereas the incalculable global minimum in the $SU(N_{c})$ theory is $\mathcal{O}(-m^{2}\Lambda^{2})$. }\label{table:phases}}

\begin{adjustbox}{width=0.75\textwidth}
\small
\begin{tabular}{|c|c|c|}

\hline
&&\\
Regime & $SU(2N_f)$ Chiral &{$SU(N_f)_L \times SU(N_f)_R$} Chiral\\
 ($N_f$ vs. $N_c$ relation)&Symmetry Breaking Vaccum &  Symmetry Breaking Vaccum  \\
 w.r.t. $Sp(N_c)$ case& in $Sp(N_c)$ theory& in $SU(N_c)$ theory\\
 &&\\
\hline\hline
&&\\
 ADS Superpotential&  Stable global minimum& Stable global  minimum\\
 $N_f\leq N_c$& 
 Residual: $Sp(2N_f)$&
 Residual: $SU (N_f )_V$\\
 &&\\
\hline
&&\\
 Quantum Modified&Stable (conjectured) global&Neither global nor local \\
 Constraint&minimum at Meson point& minima can be identified; \\
 $N_f=N_c+1$&
 Residual: $Sp(2N_f)$&baryonic runaway possible\\
 &&\\
\hline
&&\\
 s-Confinement& Stable global minimum& A stable local minimum;  \\
 $N_f=N_c+2$&
 Residual: $Sp(2N_f)$&
 Residual: $SU (N_f )_V$\\
 & &  No runaways near origin\\
 &&\\
 \hline
 &&\\
 Free Magnetic Phase&Stable global minimum&A stable local 
 minimum\\
 (Lower Part)&in Mesonic branch;&in Mesonic branch;\\
 &
 Residual: $Sp(2N_f)$&
 Residual:$SU (N_f )_ V$\\
 $N_{c} + 3 \leq N_{f} \lesssim 1.43N_c$& Mixed branch collapses &  Mixed branch runaways\\
 & to Mesonic branch&stabilized near origin\\
 &&\\
 \hline
 &&\\
 Free Magnetic Phase&Stable and calculable & Runaway to incalculable \\
 (Higher Part)&global minimum in & global minimum in \\
 &
 Mesonic branch &Dual Squark branch\\
 $1.43N_c\lesssim N_{f} \leq  \frac{3}{2} (N_{c}+1)$& Residual: $Sp(2N_f)$&with Residual\\
 &\& a local minimum in & $SU(\widetilde{N}_c) \times SU(N_{f}-\widetilde{N}_c)$  \\
 &
 Mixed q \& M branch  &\& a stable local minimum \\
 &with Residual
 &in Mesonic Branch; \\
 &$Sp(\widetilde{N}_c)\times Sp(2(N_{f}-\widetilde{N}_c))$&Residual: $SU (N_f )_ V$
 \\
 &&\\
 \hline
 &&\\
 Lower end of the &Stable and calculable & Runaway to incalculable \\
 Conformal Window&global minimum in &global minimum due to \\
 & Mesonic branch with & tachyonic dual squarks\\
 $\frac{3}{2} (N_{c}+1) < N_{f} < 3(N_{c} + 1)$&Residual $Sp(2N_f)$&with residual\\
 &\& a local minimum&$SU(\widetilde{N}_c) \times SU(N_{f}-\widetilde{N}_c)$\\
 &in Mixed q \& M branch&\& a stable local minimum \\
 &with residual &in Mesonic branch; \\
 &$Sp(\widetilde{N}_c)\times SU(2(N_{f}-\widetilde{N}_c))$&Residual: $SU (N_f )_ V$\\
 &&\\
\hline
&&\\
Upper end of the &Stable global minimum&Stable global minimum\\
Conformal Window&in Mesonic branch&in Mesonic branch\\
$\frac{3}{2} (N_{c}+1) < N_{f} < 3(N_{c} + 1)$&Residual: $Sp(2N_f)$&Residual: $SU (N_f )_ V$\\
&&\\
\hline

\end{tabular}
\end{adjustbox}
\end{center}
\end{table*}

\vspace{12pt}

\newpage

\bibliographystyle{JHEP}
\bibliography{refs2}

@article{MartinVaughn,
  title = {Two-loop renormalization group equations for soft supersymmetry-breaking couplings},
  author = {Martin, Stephen P. and Vaughn, Michael T.},
  journal = {Phys. Rev. D},
  volume = {50},
  issue = {3},
  pages = {2282--2292},
  numpages = {0},
  year = {1994},
  month = {Aug},
  publisher = {American Physical Society},
  doi = {10.1103/PhysRevD.50.2282},
  url = {https://link.aps.org/doi/10.1103/PhysRevD.50.2282}
}

@article{delima2023sconfining,
      title={On s-confining SUSY-QCD with Anomaly Mediation}, 
      author={Carlos Henrique de Lima and Daniel Stolarski},
      year={2023},
      eprint={2307.13154},
      archivePrefix={arXiv},
      primaryClass={hep-th}
}

@article{Murayama:2021xfj,
    author = "Murayama, Hitoshi",
    title = "{Some Exact Results in QCD-like Theories}",
    eprint = "2104.01179",
    archivePrefix = "arXiv",
    primaryClass = "hep-th",
    doi = "10.1103/PhysRevLett.126.251601",
    journal = "Phys. Rev. Lett.",
    volume = "126",
    number = "25",
    pages = "251601",
    year = "2021"
}

@article{Randall:1998uk,
    author = "Randall, Lisa and Sundrum, Raman",
    title = "{Out of this world supersymmetry breaking}",
    eprint = "hep-th/9810155",
    archivePrefix = "arXiv",
    reportNumber = "MIT-CTP-2788, PUPT-1815, BUHEP-98-26",
    doi = "10.1016/S0550-3213(99)00359-4",
    journal = "Nucl. Phys. B",
    volume = "557",
    pages = "79--118",
    year = "1999"
}

@article{Giudice:1998xp,
    author = "Giudice, Gian F. and Luty, Markus A. and Murayama, Hitoshi and Rattazzi, Riccardo",
    title = "{Gaugino mass without singlets}",
    eprint = "hep-ph/9810442",
    archivePrefix = "arXiv",
    reportNumber = "CERN-TH-98-337, LBNL-42419, LBL-42419, UCB-PTH-98-50, UMD-PP-99-037",
    doi = "10.1088/1126-6708/1998/12/027",
    journal = "JHEP",
    volume = "12",
    pages = "027",
    year = "1998"
}

@article{Seiberg:1994bz,
    author = "Seiberg, Nathan",
    title = "{Exact results on the space of vacua of four-dimensional SUSY gauge theories}",
    eprint = "hep-th/9402044",
    archivePrefix = "arXiv",
    reportNumber = "RU-94-18",
    doi = "10.1103/PhysRevD.49.6857",
    journal = "Phys. Rev. D",
    volume = "49",
    pages = "6857--6863",
    year = "1994"
}

@article{Seiberg:1994pq,
    author = "Seiberg, N.",
    title = "{Electric - magnetic duality in supersymmetric nonAbelian gauge theories}",
    eprint = "hep-th/9411149",
    archivePrefix = "arXiv",
    reportNumber = "RU-94-82, IASSNS-HEP-94-98",
    doi = "10.1016/0550-3213(94)00023-8",
    journal = "Nucl. Phys. B",
    volume = "435",
    pages = "129--146",
    year = "1995"
}

@article{Banks:1981nn,
    author = "Banks, Tom and Zaks, A.",
    title = "{On the Phase Structure of Vector-Like Gauge Theories with Massless Fermions}",
    reportNumber = "TAUP-944-81",
    doi = "10.1016/0550-3213(82)90035-9",
    journal = "Nucl. Phys. B",
    volume = "196",
    pages = "189--204",
    year = "1982"
}

@article{deGouvea:1998ft,
    author = "de Gouv\^ea, Andre and Friedland, Alexander and Murayama, Hitoshi",
    title = "{Seiberg duality and $e^{+} e^{-}$ experiments}",
    eprint = "hep-th/9810020",
    archivePrefix = "arXiv",
    reportNumber = "LBL-42246, LBNL-42246, UCB-PTH-98-45",
    doi = "10.1103/PhysRevD.59.105008",
    journal = "Phys. Rev. D",
    volume = "59",
    pages = "105008",
    year = "1999"
}

@article{GuideToASQCD,
  title = {Guide to anomaly-mediated supersymmetry-breaking QCD},
  author = {Cs\'aki, Csaba and Gomes, Andrew and Murayama, Hitoshi and Noether, Bea and Roy Varier, Digvijay and Telem, Ofri},
  journal = {Phys. Rev. D},
  volume = {107},
  issue = {5},
  pages = {054015},
  numpages = {10},
  year = {2023},
  month = {Mar},
  publisher = {American Physical Society},
  doi = {10.1103/PhysRevD.107.054015},
  url = {https://link.aps.org/doi/10.1103/PhysRevD.107.054015}
}

@book{Rychkov_2017,
	doi = {10.1007/978-3-319-43626-5},
  
	url = {https://doi.org/10.1007%2F978-3-319-43626-5},
  
	year = 2017,
	publisher = {Springer International Publishing},
  
	author = {Slava Rychkov},
  
	title = {{EPFL} Lectures on Conformal Field Theory in D $\geq$ 3 Dimensions}
}

@article{IntriligatorSPn,
   title={Exact superpotentials, quantum vacua and duality in supersymmetric SP(Nc) gauge theories},
   volume={353},
   ISSN={0370-2693},
   url={http://dx.doi.org/10.1016/0370-2693(95)00618-U},
   DOI={10.1016/0370-2693(95)00618-u},
   number={4},
   journal={Physics Letters B},
   publisher={Elsevier BV},
   author={Intriligator, K. and Pouliot, P.},
   year={1995},
   month=jul, pages={471–476} }

@article{Bai:2021tgl,
    author = "Bai, Yang and Stolarski, Daniel",
    title = "{Phases of confining SU(5) chiral gauge theory with three generations}",
    eprint = "2111.11214",
    archivePrefix = "arXiv",
    primaryClass = "hep-th",
    doi = "10.1007/JHEP03(2022)113",
    journal = "JHEP",
    volume = "03",
    pages = "113",
    year = "2022"
}

@article{Kondo:2021osz,
    author = "Kondo, Dan and Murayama, Hitoshi and Noether, Bea and Varier, Digvijay Roy",
    title = "{Broken conformal window}",
    eprint = "2111.09690",
    archivePrefix = "arXiv",
    primaryClass = "hep-th",
    doi = "10.1007/JHEP04(2025)152",
    journal = "JHEP",
    volume = "04",
    pages = "152",
    year = "2025"
}

@article{Intriligator:2003jj,
    author = "Intriligator, Kenneth A. and Wecht, Brian",
    title = "{The Exact superconformal R symmetry maximizes a}",
    eprint = "hep-th/0304128",
    archivePrefix = "arXiv",
    reportNumber = "UCSD-PTH-03-02",
    doi = "10.1016/S0550-3213(03)00459-0",
    journal = "Nucl. Phys. B",
    volume = "667",
    pages = "183--200",
    year = "2003"
}

@article{Caswell:1974gg,
    author = "Caswell, William E.",
    title = "{Asymptotic Behavior of Nonabelian Gauge Theories to Two Loop Order}",
    reportNumber = "PRINT-74-1058 (PRINCETON)",
    doi = "10.1103/PhysRevLett.33.244",
    journal = "Phys. Rev. Lett.",
    volume = "33",
    pages = "244",
    year = "1974"
}

\end{document}